\documentclass[floats,floatfix,showpacs,amssymb,prd,superscriptaddress,nofootinbib,aps,longbibliography,10pt]{revtex4-1}

\usepackage{graphicx}
\usepackage{epsfig, amssymb,graphicx}
\usepackage{epstopdf}
\usepackage{amsmath, amsfonts}
\usepackage{subfigure}
\usepackage{caption}
\usepackage{bm} 
\usepackage[linktocpage]{hyperref}

\graphicspath{ {./figures/} }

\begin{document}

\title{\large Tensor Modes in Bigravity: Primordial to Present}

\author{Matthew C. Johnson}\email{mjohnson@perimeterinstitute.ca}
\affiliation{Department of Physics and Astronomy, York University
Toronto, On, M3J 1P3, Canada}
\affiliation{Perimeter Institute for Theoretical Physics Waterloo, Ontario N2J 2W9, Canada}

\author{Alexandra Terrana}\email{aterrana@perimeterinstitute.ca}
\affiliation{Department of Physics and Astronomy, York University
Toronto, On, M3J 1P3, Canada}
\affiliation{Perimeter Institute for Theoretical Physics Waterloo, Ontario N2J 2W9, Canada}

\date{\today} 

\begin{abstract}
Massive bigravity, a theoretically consistent modification of general relativity with an additional dynamical rank two tensor, successfully describes the observed accelerated expansion of the Universe without a cosmological constant. Recent analyses of perturbations around a cosmological background have revealed power law instabilities in both scalar and tensor perturbations, motivating an analysis of the initial conditions, evolution, and cosmological observables to determine the  viability of these theories. In this paper we focus on the tensor sector, and study a primordial stochastic gravitational wave background in massive bigravity. The phenomenology can differ from standard General Relativity due to non-trivial mixing between the two linearized tensor fluctuations in the theory, only one of which couples to matter. We study perturbations about two classes of cosmological solutions in bigravity, computing the tensor contribution to the temperature anisotropies in the Cosmic Microwave Background radiation and the present stochastic gravitational wave background. The result is strongly dependent on the choice of cosmological background and initial conditions. One class of background solution generically displaying tremendous growth in the amplitude of large-wavelength gravitational waves, while the other remains observationally indistinguishable from standard General Relativity for a wide variety of initial conditions. We analyze the initial conditions for tensor modes expected in an inflationary cosmology, finding again that there is a strong dependence on the assumed background. For one choice of background, the semi-classical theory is beyond the perturbative regime. For the other choice, inflation generically yields initial conditions that, when evolved, give rise to a stochastic background observationally indistinguishable from standard General Relativity.  
\end{abstract}

\pacs{ 
04.50.Kd,
04.30.-w	
04.80.-y	
98.80.Es	
95.36.+x	
} 

\maketitle


\section{Introduction} \label{sec:intro}

Extending general relativity (GR) by giving a mass to the graviton has proven to be a difficult endeavor, dating back to first attempts by Fierz and Pauli in 1939 \cite{Fierz:1939ix}. As a result of recent work in constructing a fully non-linear ghost-free theory, dRGT massive gravity \cite{deRham:2010kj}, interest in this field has been on the rise (see \cite{deRham:2014zqa} for a review). An important yet curious feature of this theory is the need to introduce a fixed non-dynamical tensor field,  $f_{\mu\nu}$, in addition to the metric $g_{\mu\nu}$ describing spacetime. Promoting this tensor to a dynamical field, a ghost-free theory of \emph{massive bigravity} \cite{Hassan:2012aa} has recently been introduced. With two dynamical metrics, the choice of matter coupling is a non-trivial issue~\cite{de-Rham:2014aa}; we will consider couplings of matter to one metric only. In this scenario, one metric $g_{\mu\nu}$ describes our spacetime, and another one, $f_{\mu\nu}$, is part of a ``dark" gravitational sector.\\

Much of the interest in massive gravity is motivated by the puzzle of cosmic acceleration and dark energy. In the standard cosmological model, this accelerated expansion is assumed to be due to a cosmological constant. However, the extreme fine tuning of the cosmological constant has led physicists to pursue alternative explanations for cosmic acceleration. Viable homogenous and isotropic cosmological solutions exist in bigravity which can describe our universe, including acceleration, without a cosmological constant \cite{Koennig:2014ab,Akrami:2013ab,von-Strauss:2012aa,Tamanini:2014aa,Comelli:2012ab,Volkov:2012aa,Solomon:2014aa}. These are usually referred to as \emph{self-accelerating} background solutions. Beyond the background level, investigations  have recently been underway to analyze perturbations in bigravity\cite{Comelli:2012aa,Koennig:2014aa,Berg:2012aa,Solomon:2014aa,Lagos:2014aa,Cusin:2014aa}. These analyses reveal that there is a particular class of stable solutions, but all others are plagued by an exponential instability in the scalar sector in the early universe \cite{Solomon:2014aa}. \\

The transverse traceless fluctuations of each of the two dynamical metrics in bigravity interact, altering the propagation of gravitational waves as compared to GR. The most general set of linearized equations of motion for the visible sector tensor modes $h_g$ and the dark sector tensor modes $h_f$ is given by
\begin{equation}
\mathbf{D}^2 \cdot \mathbf{h} + \mathbf{m}^2 (\mathbf{x},\tau) \cdot \mathbf{h} = 0
\end{equation}
where 
\begin{equation}
\mathbf{h} = \left(\begin{array}{c}
h_f \\
h_g
\end{array}
\right), \ \ 
\mathbf{D}^2 \equiv \left(\begin{array}{c c}
\nabla_f^2 \ \ 0 \\
0 \ \ \nabla_g^2
\end{array}\right), \ \ 
\mathbf{m}^2 (\mathbf{x},\tau)
\equiv \left(\begin{array}{c c}
m_f^2 (\mathbf{x}, \tau) \ \ m_{fg}^2  (\mathbf{x}, \tau) \\
m_{gf}^2  (\mathbf{x}, \tau)  \ \ m_g^2  (\mathbf{x}, \tau) 
\end{array}\right),
\end{equation}
where $\nabla_f^2$ is the covariant derivative defined with respect to $f_{\mu \nu}$ and $\nabla_g^2$ is the covariant derivative defined with respect to $g_{\mu \nu}$. There is no off-diagonal term in the differential operator $\mathbf{D}^2$ due to the absence of consistent derivative couplings between the two metrics. The mass matrix $\mathbf{m}^2$ is in general not diagonal or symmetric. The differential operator $\mathbf{D}^2$ and mass matrix $\mathbf{m}^2$ are generally not simultaneously diagonalizable, leading to mixing between the visible and dark sector tensors. Because the visible and dark sector tensor fluctuations evolve in a different (spacetime dependent) background and possess a different (spacetime dependent) mass, their propagation speeds can in general be different and spacetime dependent. The modifications in the propagation speed of gravitational waves and mixing between the visible and dark sector tensors generically present in bigravity can have important implications for both astrophysical~\cite{DeFelice:2013nba,Brito:2013wya,Saltas:2014dha,Berezhiani:2007aa} and primordial~\cite{Raveri:2014eea,Xu:2014uba,Amendola:2014wma,Pettorino:2014bka,Cai:2015dta,Amendola:2015tua} gravitational waves. \\

In this paper, we consider a stochastic background of primordial gravitational waves in massive bigravity, studying the impact on primordial and present day gravitational wave observables by computing the tensor contribution to the temperature anisotropies of the Cosmic Microwave Background radiation (CMB) and the power spectrum of the present day stochastic gravitational wave background. We examine whether these observables can be used to distinguish between bigravity and GR, serving as a possible test for gravity on cosmological scales. We consider two classes of cosmological background solutions for the dark sector metric $f_{\mu \nu}$. For one class, the tensor perturbations match those of GR very closely. However, for the second class of background solution, a power-law instability in the tensor sector leads to a strong growth in the amplitude of gravitational waves at late times. In the absence of special initial conditions, these solutions are in conflict with observations. We argue that these conditions may be naturally produced in the standard inflationary picture. In addition, we explore to what extent this fine tuning is present in solutions that are not self accelerating. \\

During the preparation of this manuscript, the analysis of Cusin et. al.~\cite{Cusin:2014aa} appeared; additionally the analysis of Amendola et. al.~\cite{Amendola:2015tua} appeared. There is significant overlap in our discussion of the evolution of perturbations. However, we make substantially different assumptions about the initial conditions, which we motivate from a detailed discussion of bigravity in the context of inflationary cosmology. This analysis addresses an outstanding question regarding the viability of massive bigravity first outlined in Cusin et. al. We explicitly comment on the relevant differences in our assumptions throughout the manuscript. \\

The paper is organized as follows: In section \ref{sec:bigravity} we give the bigravity background equations and the two classes of solutions under consideration. Section \ref{sec:tensors} analyzes the gravitational waves in each branch. In the following sections we compute gravitational wave observables: section \ref{sec:CMB} gives the CMB Tensor Power Spectrum in bigravity, and section \ref{sec:OmegaGW} gives the present day stochastic gravitational wave background. Section \ref{sec:initialconds} explores the initial conditions as predicted by inflation, followed by a discussion in \ref{sec:conclusion}.

\section{Bigravity Cosmology} \label{sec:bigravity}

We will consider the massive bigravity model proposed by Hassan and Rosen \cite{Hassan:2012aa}:

\begin{equation} \label{eq:action}
	S =  -\frac{M_g^2}{2} \int{d^4x \sqrt{-g}R(g)} - \frac{M_f^2}{2} \int{d^4x \sqrt{-\tilde{f}}R(\tilde{f})} + m^2 M_g^2 \int{d^4x \sqrt{-g} \sum_{n=0}^{4}{\tilde{\beta}_n e_n(\sqrt{g^{-1}\tilde{f}})}} + S_\text{matter}
\end{equation}
where the two dynamical metrics are $g_{\mu\nu}$ and $\tilde{f}_{\mu\nu}$ and their respective Planck masses are $M_g$ and $M_f$, while $m$ is the mass scale associated with the graviton mass matrix. The interaction term between $g$ and $\tilde{f}$ contains a linear combination of the symmetric polynomials $e_n(X)$ which are defined by
\begin{align}
	e_0(X) = & 1 \\
	e_1(X) = & [X] \\
	e_2(X) = & \frac{1}{2}([X]^2 - [X^2]) \\
	e_3(X) = & \frac{1}{6}([X]^3 - 3[X][X^2] + 2[X^3]) \\
	e_4(X) = & \det(X)
\end{align}
given in terms of the tensor $X^\mu_\nu = \sqrt{g^{\mu\alpha}\tilde{f}_{\alpha\nu}}$ where square brackets denote the trace. The dimensionless coefficients $\tilde{\beta}_n$ are free parameters of the theory.  Further, we will consider a singly-coupled theory in which $S_\text{matter}$ contains only couplings of matter to $g_{\mu\nu}$. In this case, $g_{\mu\nu}$ is considered the standard physical metric while $\tilde{f}_{\mu\nu}$ is a new dynamical tensor field. This singly-coupled theory was shown to be free of the Boulware-Deser ghost \cite{de-Rham:2014aa}. The resulting quantum corrections (at one loop) are nothing other than the standard cosmological constant which does not detune the special structure of the graviton potential and is thus harmless \cite{de-Rham:2014aa}. 
By rescaling the dark metric  and the free parameters as follows,
\begin{equation}
f_{\mu\nu}=(M_f/M_g)^2\tilde{f}_{\mu\nu} \ \ \ \ \ \ \beta^*_n=(M_g/M_f)^n\tilde{\beta}_n \  ,
\end{equation} 
 we can rewrite the action so that the redundant scale $M_f$ is absent: 
\begin{equation} \label{eq:action2}
	S =  M_g^2 \left[-\frac{1}{2} \int{d^4x \sqrt{-g}R(g)} - \frac{1}{2} \int{d^4x \sqrt{-f}R(f)} + m^2 \int{d^4x \sqrt{-g} \sum_{n=0}^{4}{\beta^*_n e_n(\sqrt{g^{-1}f})}} + \frac{S_\text{matter}}{M_g^2} \right]
\end{equation}
	It has been shown that the cosmological expansion and spherically-symmetric solutions to this theory give viable alternatives to general relativity \cite{Koennig:2014ab,Akrami:2013ab,von-Strauss:2012aa,Tamanini:2014aa,Comelli:2012ab,Volkov:2012aa,Solomon:2014aa,Comelli:2012ac}, at the background level. Let us briefly examine the background equations of motion and solutions.\\
	
	We start by making an FRW ansatz for the metrics $g_{\mu\nu}$ and $f_{\mu\nu}$:
\begin{align} \label{eq:metricg}
	ds_g^2 =& a^2(\tau)(-d\tau^2+dx_idx^i) \\ \label{eq:metricf}
	ds_f^2 =& b^2(\tau)[-c^2(\tau)d\tau^2 + dx_idx^i] 
\end{align}
where $\tau$ represents conformal time,  $a$ and $b$ are the scale factors corresponding to the $g$ and $f$ metric respectively, and $c$ is the lapse function for the $f$ metric. Note that the assumption that $f_{\mu\nu}$ is FRW is not the most general choice for the metric,  and it could have more complicated dynamics. \\

The Bianchi identities imply the following relation:
\begin{equation} \label{eq:lapse}
	c = \frac{\hat{\mathcal{H}}_f}{\hat{\mathcal{H}}} = \frac{\dot{b}a}{\dot{a}b} = 1 + \frac{\dot{r}}{r\hat{\mathcal{H}}}.
\end{equation}
where $\hat{\mathcal{H}}=\dot{a}/a$ is the conformal Hubble function for the $g$ metric, $\hat{\mathcal{H}}_f=\dot{b}/b$ is the conformal Hubble function for the $f$ metric, and $r = b/a$ the ratio of scale factors. For simplicity we perform the following rescaling
\begin{equation} \label{eq:betas}
	\beta_n = \frac{m^2}{H_0^2}\beta_n^* \ \ \ \ \ \ \bar{\rho}=\frac{\rho}{M_g^2H_0^2} \ \ \ \ \ \ \mathcal{H}=\frac{\hat{\mathcal{H}}}{H_0}
\end{equation}
where $\bar{\rho}=\bar{\rho}_m+\bar{\rho}_r$ is the dimensionless energy density of all matter and radiation components, and we measure all times and lengths in terms of $H_0$. With these definitions in hand, variation of the action with respect to $g_{\mu\nu}$, and inserting \eqref{eq:metricg} and \eqref{eq:metricf}, gives the Friedmann equations
\begin{eqnarray} \label{eq:eomg1}
	3\mathcal{H}^2 &=& a^2\left(\bar{\rho} + \bar{\rho}_\text{mg}\right) \\ \label{eq:eomg2}
	2\dot{\mathcal{H}}+\mathcal{H}^2 &=& a^2\bar{\rho}+\frac{a^3}{3}\frac{d\bar{\rho}}{da} +  a^2\left( \beta_0 + \beta_1 r (2+c)+\beta_2 r^2(1+2c)+\beta_3r^3c)  \right) \\
	\text{where} & \ &\ \bar{\rho}_\text{mg} \equiv \left( \beta_0 + 3\beta_1r + 3\beta_2 r^2 + \beta_3 r^3 \right)
\end{eqnarray}
where $\bar{\rho}_\text{mg}$ is an effective massive-gravity energy density. We also have the background equations for the $f$ metric:
\begin{eqnarray} \label{eq:eomf1}
	3\mathcal{H}^2 &=& \frac{a^2}{r}\left(\beta_1 + 3\beta_2r + 3\beta_3 r^2 + \beta_4 r^3\right) \\ \label{eq:eomf2}
	2\dot{\mathcal{H}} + \mathcal{H}^2c&=& \frac{a^2}{r}\left( \beta_1 + \beta_2 r (2+c) + \beta_3 r^2(1+2c)+\beta_4r^3c \right)
\end{eqnarray}
The energy densities follow the usual conservation laws
\begin{equation} \label{eq:conservation}
	\dot{\bar{\rho}}_m + 3\mathcal{H}\bar{\rho}_m = 0 \ \ \ \ \text{and} \ \ \ \ \dot{\bar{\rho}}_r + 4\mathcal{H}\bar{\rho}_r = 0 \ ,
\end{equation}
giving rise to their solutions in terms of $a$ and the present day density parameters $\Omega_i^0=\rho_i^0/(3H_0^2M_g^2) = \bar{\rho}_i^0/3$: 
\begin{equation} \label{eq:rho}
	\bar{\rho}_m = \frac{\Omega_m^0}{3a^3}\ \ \ \ \text{and} \ \ \ \ \ \bar{\rho}_r =\frac{\Omega_r^0}{3a^4} .
\end{equation}

The above equations can be combined to form convenient equations for the dynamic variables $r$ and $a$:

\begin{eqnarray} \label{eq:quarticr}
	  0 &=&\bar{\rho}_m + \bar{\rho}_r -\frac{1}{r}\left( \beta_1 + 3\beta_2r + 3\beta_3 r^2 + \beta_4 r^3 \right) +\beta_0 + 3\beta_1r + 3\beta_2 r^2 + \beta_3 r^3 \\ \label{eq:drda}
	  \frac{1}{r}\frac{dr}{da} &=& -\frac{3}{a}\frac{\beta_3r^4+(3\beta_2-\beta_4)r^3+3(\beta_1-\beta_3)r^2+(\beta_0-3\beta_2)r-\beta_1-\Omega_r^0a^{-4}r}{3\beta_3r^4+2(3\beta_2-\beta_4)r^3+3(\beta_1-\beta_3)r^2+\beta_1}
\end{eqnarray}

The procedure now is as follows:
\begin{enumerate}
	\item Fix the free parameters: $\Omega_m^0,\ \Omega_r^0,\ \beta_0,\ \beta_1,\ \beta_2,\ \beta_3,\ \beta_4$.
	\item Use \eqref{eq:quarticr} evaluated today to find $r_0$, the present value of $r$.
	\item Starting with initial condition $r_0$, evolve \eqref{eq:drda} to find $r(a)$ for all time. 
	\item Using $r(a)$, solve for the Hubble parameter $\mathcal{H}(a)$ using \eqref{eq:eomf1}.
	\item It is now possible to find $a(\tau)$ using $\mathcal{H} = \dot{a}/a$, and thus can also find $r(\tau)$ and $\mathcal{H}(\tau)$ and $c(\tau)$
\end{enumerate}

\begin{figure}
		\begin{center}
			\subfigure{\includegraphics[width=7.5cm]{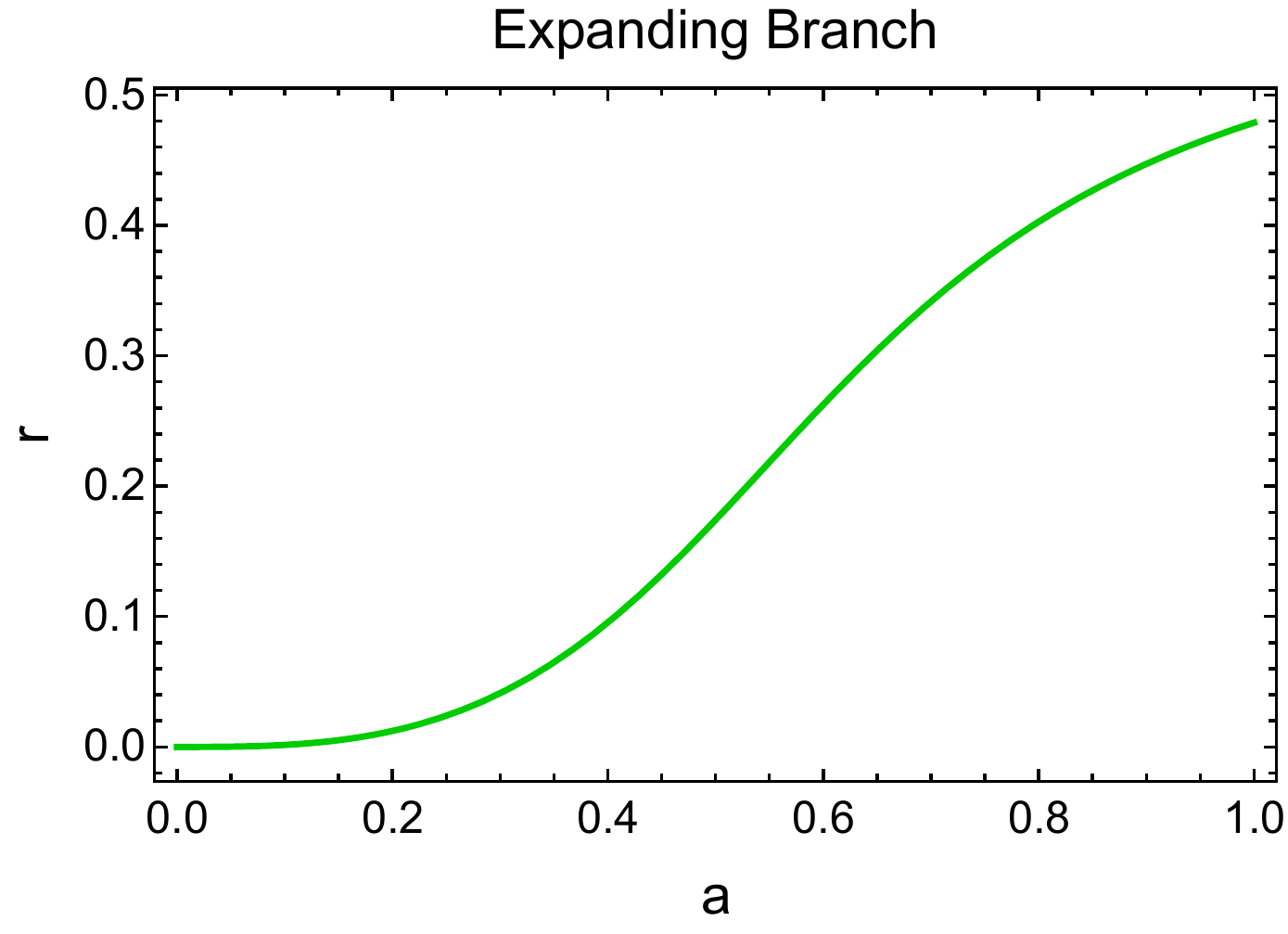}}
			\hspace{1cm}
			\subfigure{\includegraphics[width=7.5cm]{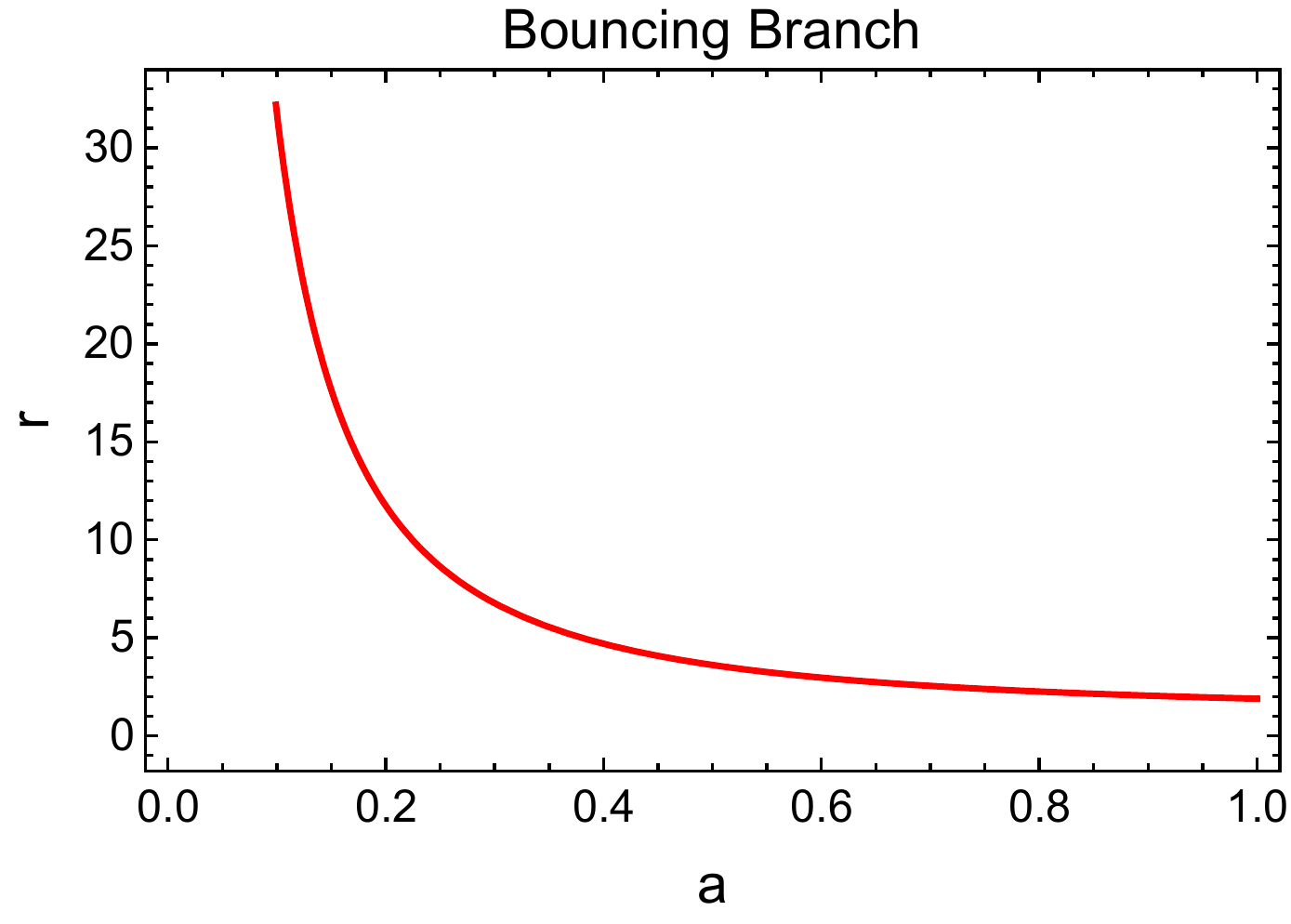}}
			\captionsetup{singlelinecheck=off,font=footnotesize}
			\caption{The evolution of the ratio of scale factors $r=b/a$ for the both branches described in the text.}
			\label{fig:ra}
		\end{center}
\end{figure}

In this work we will focus on two distinct types of background solutions, following the notation in \cite{Lagos:2014aa}. A plot of $r(a)$ in each branch is shown in FIG. \ref{fig:ra}.

\begin{enumerate}

	\item \textbf{The Expanding Branch}: In this branch of solution, both metrics $g$ and $f$ expand with time. This is also known as the finite branch since the ratio $r=b/a$ evolves from zero at early times to a finite value. Within this branch, there is a proposed minimal model \cite{Koennig:2014aav1,Koennig:2014ab} in which only $\beta_1 \neq 0$. At the background level, it was shown that this theory can be compatible with expansion histories, but remains distinct from GR with testably different observables \cite{Koennig:2014aav1,Koennig:2014ab}. The issue with this branch is an exponential perturbative instability in the scalar sector, previously noted in the literature \cite{Comelli:2012aa,Koennig:2014aa,Konnig:2014aa,Lagos:2014aa}. We therefore focus on this branch as an example of how phenomenologically different the results for tensors can be under different assumptions about the background solution. We will work within the minimal model, fixing $\beta_1=1.38$, which was found from a best fit analysis at the background level in \cite{Koennig:2014ab}. In this case, an analytic solution exists for $r(a)$ and is given by
\begin{equation}
	r(a)=\frac{-3a\Omega_m^0-3\Omega_r^0+\sqrt{3(4a^8\beta_1^2+3a{\Omega_m^0}^2+6a\Omega_m^0\Omega_r^0+3{\Omega_r^0}^2)}}{6a^4\beta_1}
\end{equation}
	
\item \textbf{The Bouncing Branch}:	This is a more exotic option in which the physical metric $g$ expands in time while the dark sector metric $f$ experiences a bounce. At the bounce point, $f_{00}=0$, thus, $ f^{-1}_{\mu\nu}$ diverges, but there is no divergence in the physical sector \cite{Lagos:2014aa,Gratia:2013aa,Gratia:2014aa}. In addition, well-defined and stable solutions for the background and linear perturbations exist through this point \cite{Lagos:2014aa} indicating that this divergence is likely of a mathematical rather than physical nature. The perturbative instability of the expanding branch is absent in this sector. 
This branch is also called the infinite branch since $r$ evolves from infinity at early times to a finite value at late times. Here we are required to set $\beta_0=\beta_2=\beta_3=0$, and for the remaining parameters, we use the best-fit values $\beta_1=0.48$ and $\beta_4=0.94$ found by fitting growth histories and type Ia supernovae \cite{Konnig:2014aa}.

\end{enumerate}
	
\section{Tensor Perturbations} \label{sec:tensors}

Here we give the equations of motion for the transverse traceless tensor modes $h^{\text{TT}}_{g,ij}$ and $h^{\text{TT}}_{f,ij}$ corresponding to the metrics $g_{\mu\nu}$ and $f_{\mu\nu}$ respectively. To compute observables, we will be most interested in the perturbations corresponding to the physical metric, $h^{\text{TT}}_{g,ij}$, since these are the ones that couple to matter. The tensor equations of motion were first analyzed in \cite{Comelli:2012aa}. A thorough analysis of scalar, vector, and tensor perturbations was performed in \cite{Lagos:2014aa}. In addition, further analysis of tensor perturbations appeared in Refs.~\cite{Cusin:2014aa} and~\cite{Amendola:2015tua}. The tensor perturbation equations of motion in momentum space are
\begin{align} \label{eq:tensorg}
	& \ddot{h}_g + 2\mathcal{H} \dot{h}_g +k^2 h_g + a^2 \lambda (h_g - h_f) = 0 \\ \label{eq:tensorf}
	& \ddot{h}_f + \left[ 2\left( \mathcal{H} + \frac{\dot{r}}{r} \right) - \frac{\dot{c}}{c} \right] \dot{h}_f + c^2k^2 h_f - \frac{a^2 \lambda c}{r^2} (h_g - h_f) = 0
\end{align}
where superscripts and subscript indices have been dropped for simplicity. In addition, the time-dependent function $\lambda$ is defined as
\begin{equation} \label{eq:lambda}
	\lambda = \beta_3 c r^3 + \beta_2(c+1)r^2 + \beta_1r
\end{equation}
which simplifies to $	\lambda=  \beta_1r$ in either branch under consideration. These equations are satisfied separately for each polarization; the polarizations do not mix. \\

Initial conditions for $h_g$ and $h_f$ at some initial time $\tau_i$ are required to obtain solutions. In the absence of a theory of initial conditions, we should consider general initial data $h_{(g,f)}(\tau_i)$ and  $\dot{h}_{(g,f)}(\tau_i)$. However, for standard inflationary cosmology in GR, tensor modes freeze in once their physical wavelength becomes comparable to the primordial horizon size, motivating $\dot{h}_{g}(\tau_i)=0$. In Sec.~\ref{sec:initialconds}, we compute the initial conditions expected for inflationary cosmology in the context of bigravity, finding that $\dot{h}_{(g,f)}(\tau_i)=0$ is an appropriate choice. Note that this assumption differs from Cusin et. al., who consider initial data with $\dot{h}_{f}(\tau_i) \neq 0$. This was motivated by the presence of a growing mode for $h_{f}$, which if excited, would dominate the evolution. Our choice of initial data initially sets this growing mode to zero, and as shown below, this leads to a different growth history for $h_{f}$ at late times.

\subsubsection{Expanding Branch}

\begin{figure}
			\subfigure{\includegraphics[width=7.5cm]{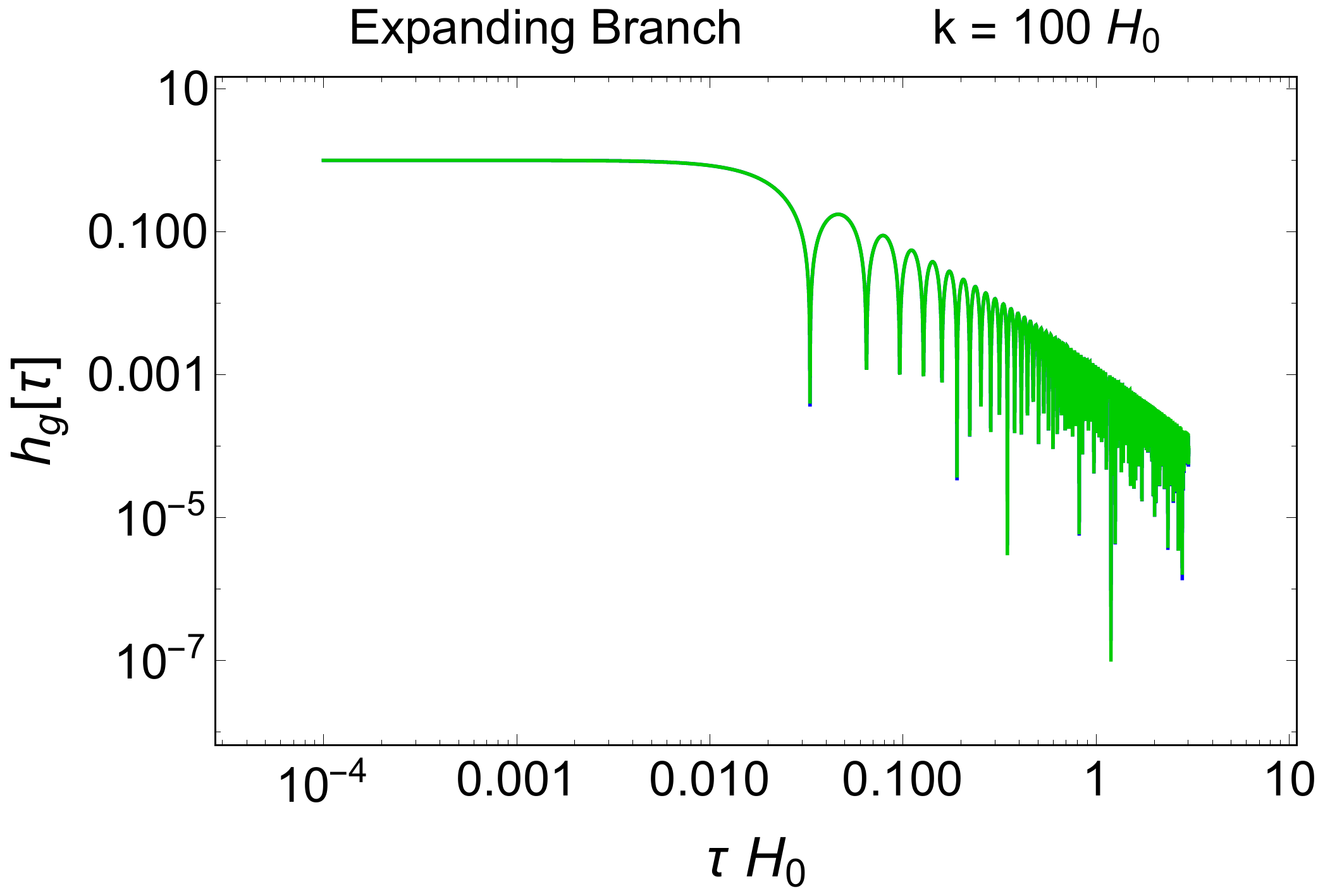}}
			\hspace{1cm}
			\subfigure{\includegraphics[width=7.5cm]{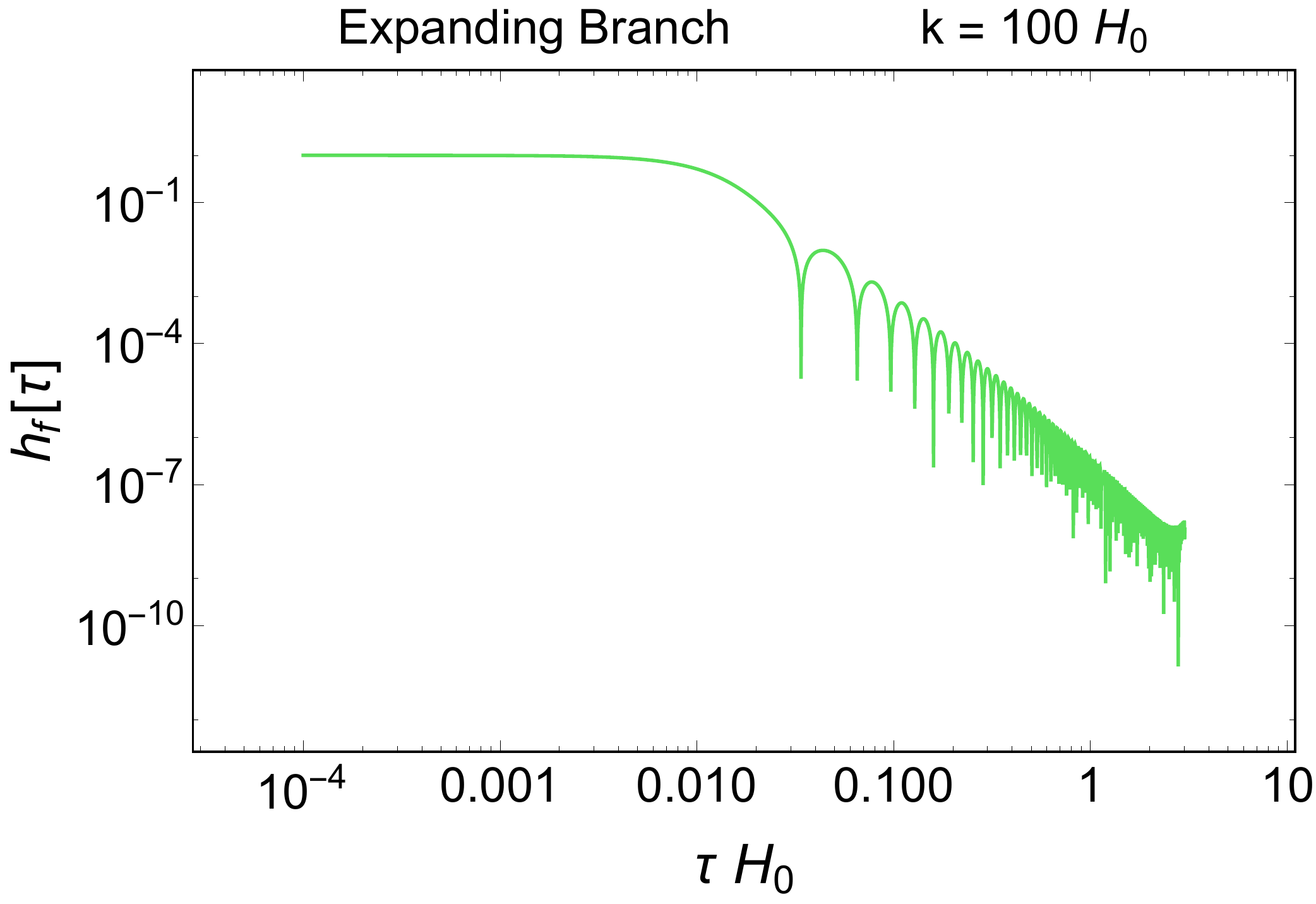}}\\
			\subfigure{\includegraphics[width=7.5cm]{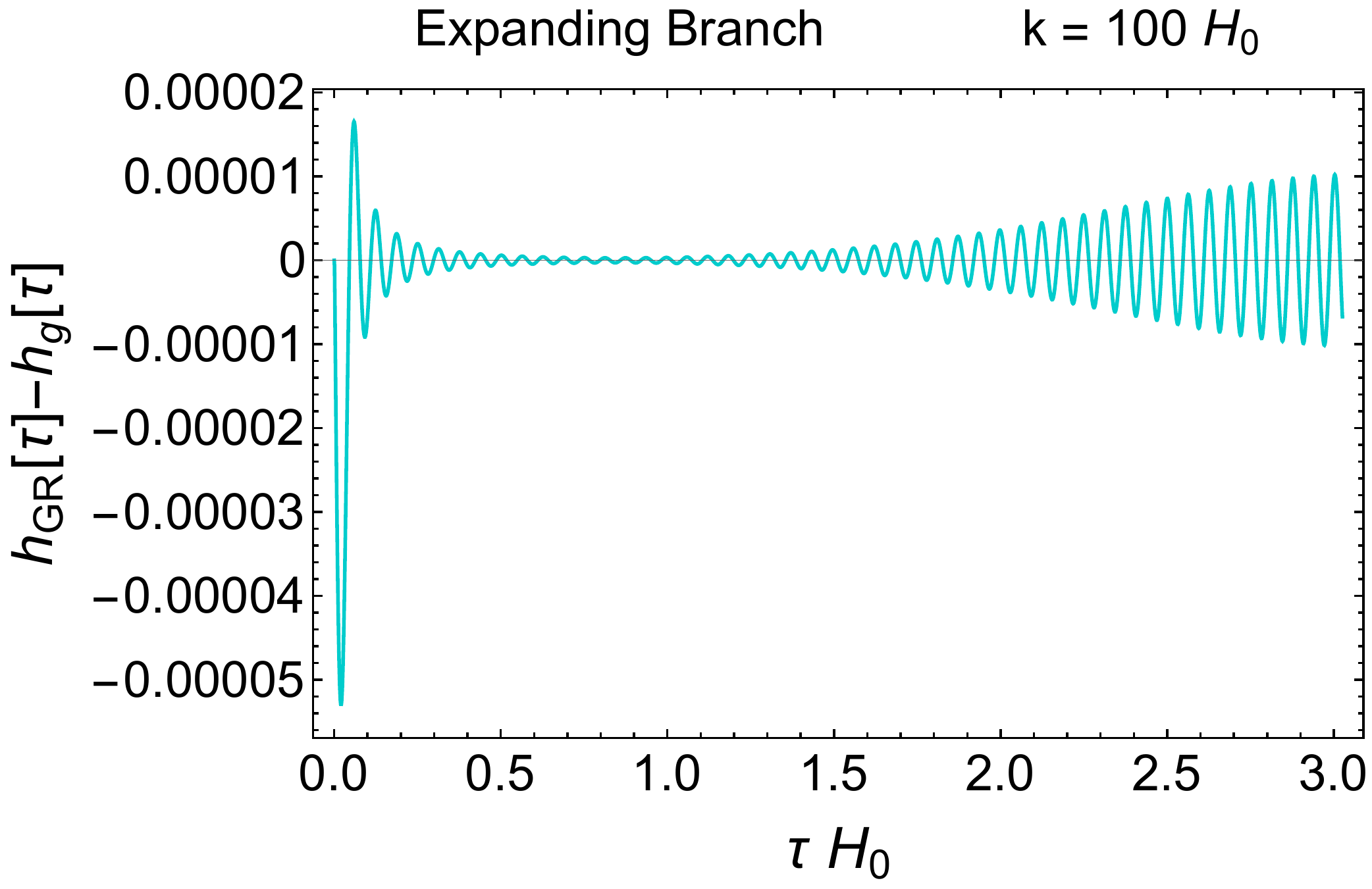}}
			\hspace{1cm}
			\subfigure{\includegraphics[width=7.5cm]{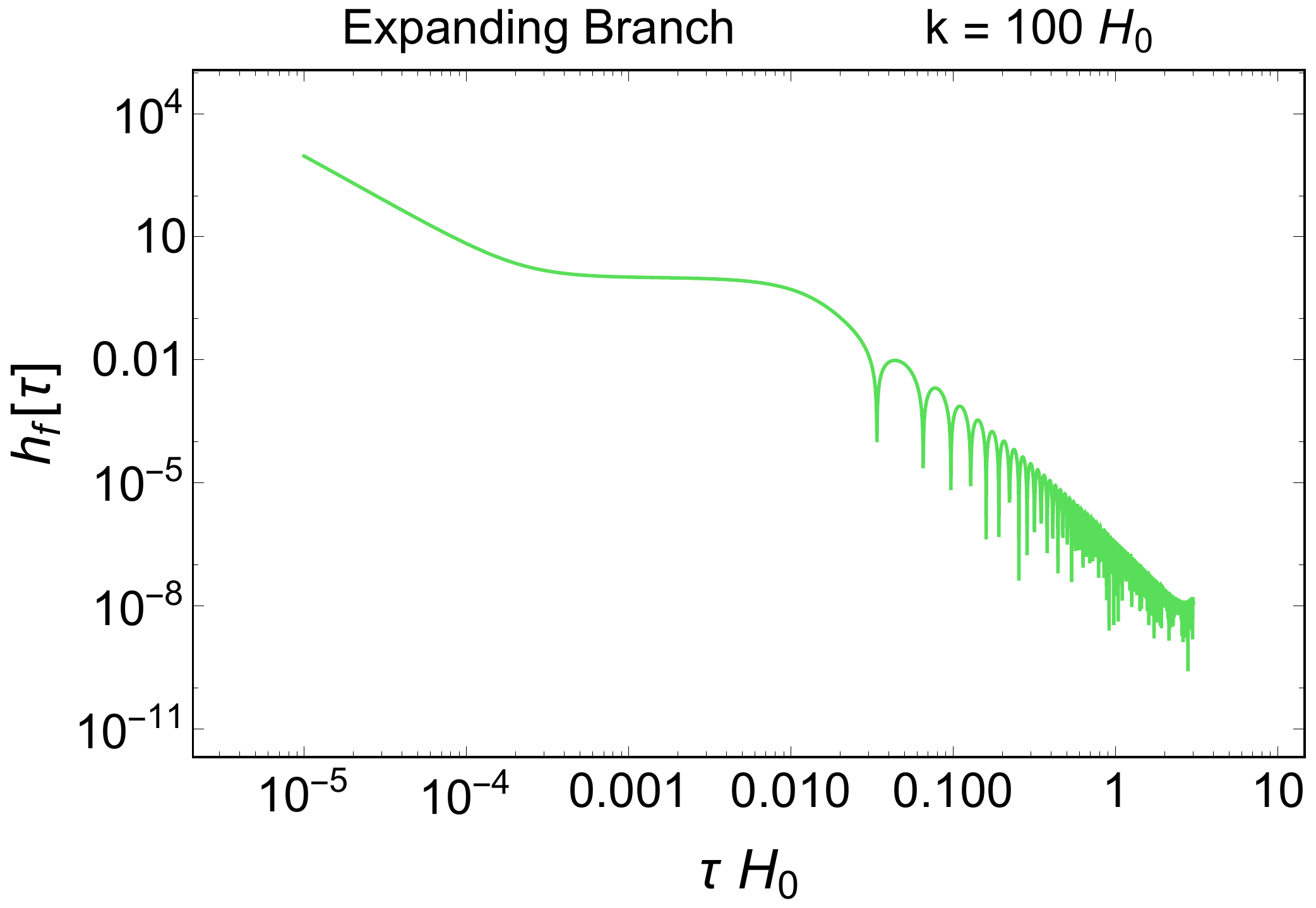}}
			\captionsetup{justification=raggedright,singlelinecheck=off,font=footnotesize}
			\caption{\textbf{Top left}: the solution for $h_g(\tau)$ at $k=100 H_0$ for the expanding branch (green) versus GR (blue). The two solutions are essentially indistinguishable. \textbf{Top right}: the solution for $h_f(\tau)$ at $k=100 H_0$ for the expanding branch. \textbf{Bottom left}: The difference between the expanding branch solution and the standard GR solution, shown to be less than $5\times 10^{-5}$ for $k=100 H_0$. \textbf{Bottom right}: the solution for $h_f(\tau)$ at $k=100 H_0$ for the expanding branch, with $h_f(\tau_i)/h_g(\tau_i)=10^6$. Even in this case of $h_f(\tau_i)\gg h_g(\tau_i)$, the decay of $h_f$ is so fast that it does not cause any alteration to the physical tensor perturbations.}
			\label{fig:tex100}
\end{figure}

In the expanding branch, the factor $\left[ 2\left( \mathcal{H} + \frac{\dot{r}}{r} \right) - \frac{\dot{c}}{c} \right]$ is always positive, and causes significant damping of the dark sector tensor perturbation $h_f$. In addition, the factor $a^2\lambda$ is small in this branch, $a^2\lambda < 0.3$ for all time. Therefore, unless the initial amplitude of $h_f$ is very large compared with $h_g$, the mixing term $a^2 \lambda (h_g - h_f)$ does not significantly alter the behaviour of the physical tensor perturbation $h_g$. For equal initial amplitude, our numerical solution for the tensor modes in this branch match closely with those of pure GR. This was confirmed for modes ranging from $k=0.1H_0$ to $k=10^5H_0$. Refer to FIG. \ref{fig:tex100} for the numerical solutions of this branch as compared to the standard GR gravitational waves with $\tau_i = 10^{-6} H_0^{-1}$.~\footnote{Although this is choice for $\tau_i$, corresponding to reheat temperature $T_i=T_\text{eq}[a_\text{eq}/a(\tau_i)]= T_0 (a_0/a_\text{eq})^2[a_\text{eq}/a(\tau_i)]=0.07$ GeV, is not entirely plausible, it is still deep within the radiation era and is a practical choice for our numerical analysis. We will be able to extrapolate to earlier $\tau_i$ when necessary using scaling properties of the solutions defined in \eqref{eq:growth}.} 
\\

If the relative amplitude of the dark sector tensor mode was decreased, $h_f(\tau_i) < h_g(\tau_i)$, this would only drive the physical tensor modes closer to those of GR. But if $h_f$ had a much higher initial amplitude relative to $h_g$, the mixing term $a^2\lambda h_f$ in \eqref{eq:tensorf} can become dominant for some time. However, even if we set $h_f(\tau_i) \gg h_g(\tau_i)$, $h_f$ decays so dramatically that there is very little effect on $h_g$. The bottom right plot in FIG. \ref{fig:tex100} shows how the dark sector perturbation decays very quickly, even when starting with a much larger amplitude. \\

\subsubsection{Bouncing Branch} \label{sub:tensorsbb}

\begin{figure}
		\begin{center}
			\subfigure{\includegraphics[width=7.5cm]{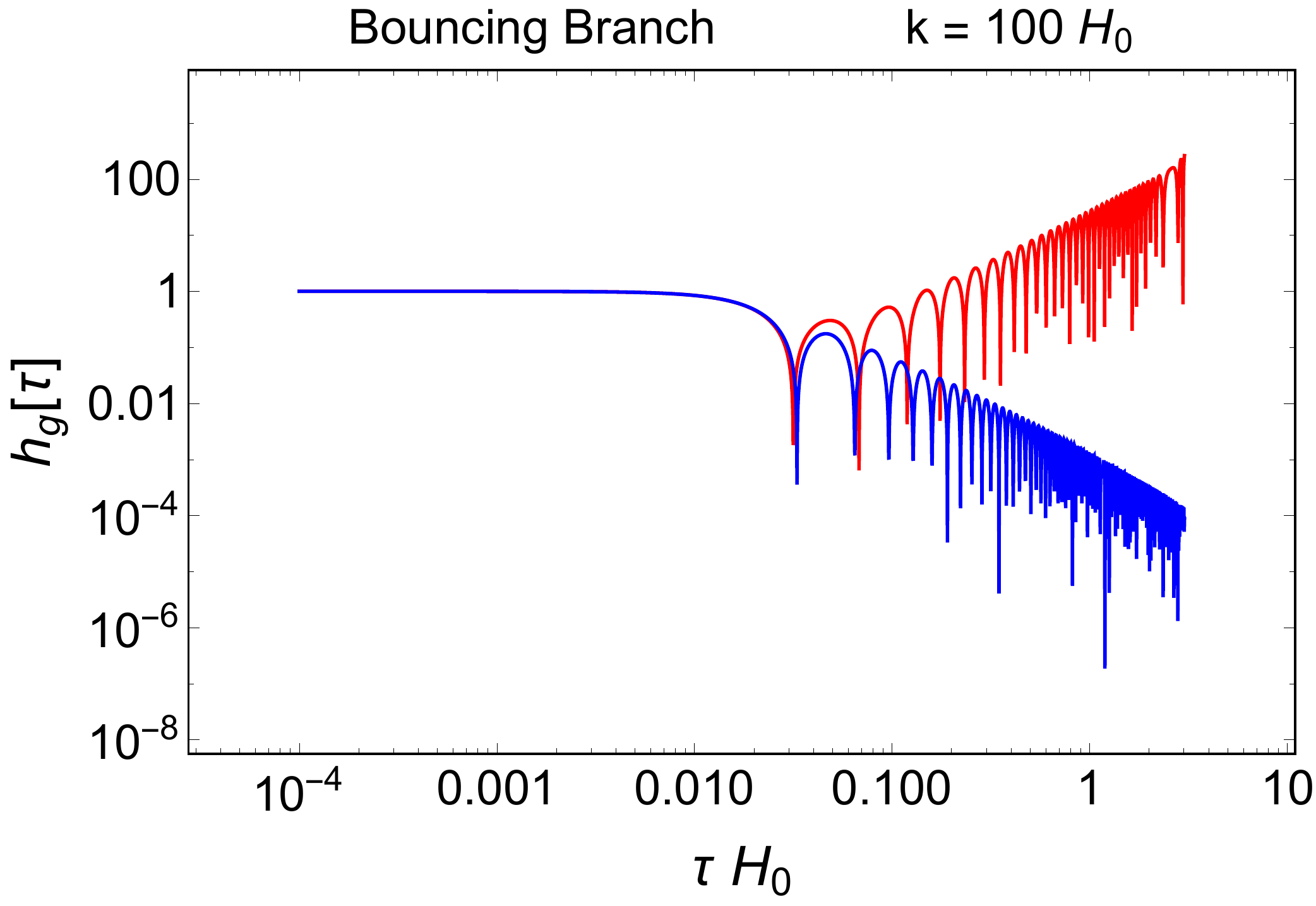}}
			\hspace{1cm}
			\subfigure{\includegraphics[width=7.5cm]{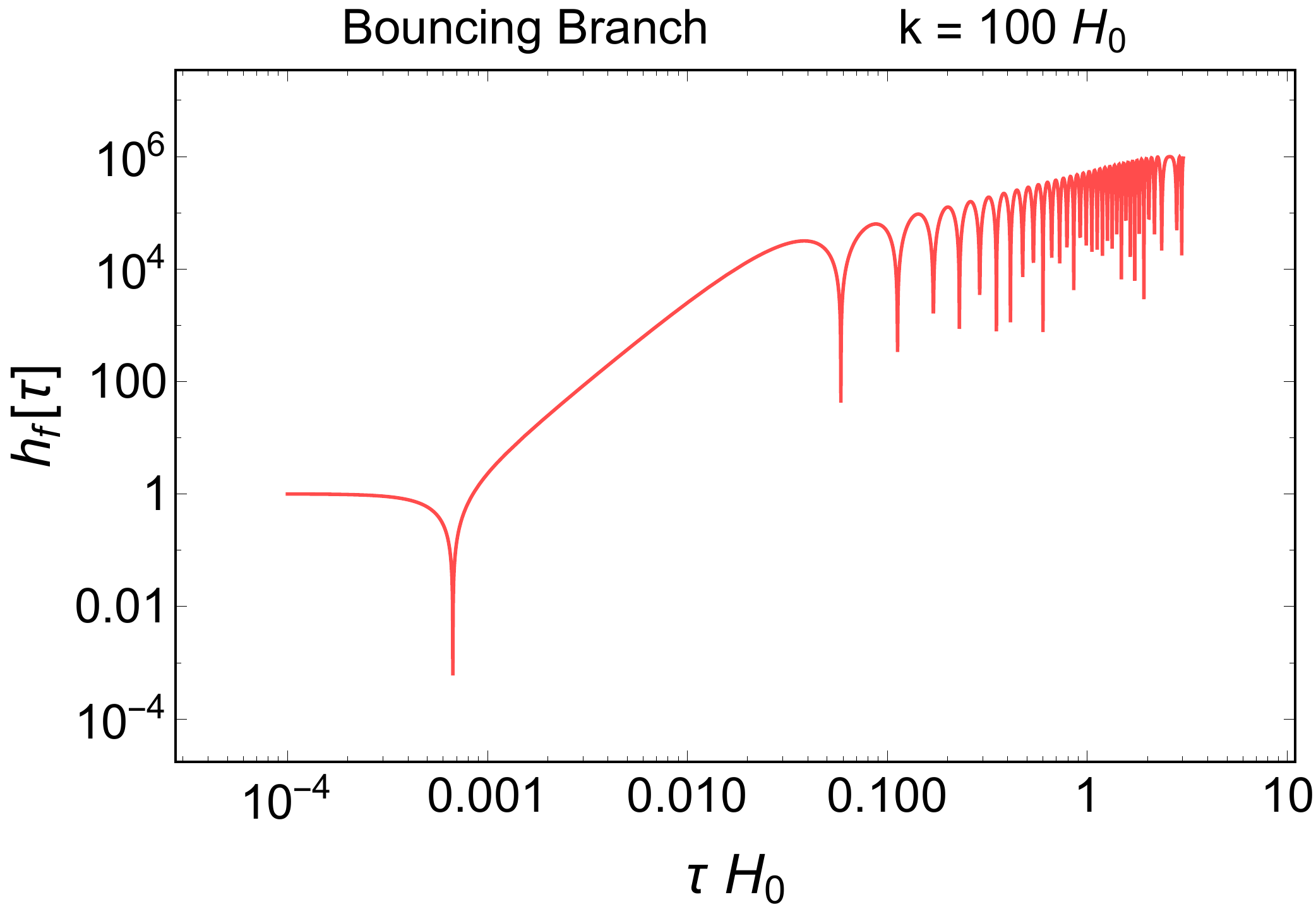}}
			\captionsetup{justification=raggedright,singlelinecheck=off,font=footnotesize}
			\caption{\textbf{Top left}: the solution for $h_g(\tau)$ at $k=100 H_0$ for the bouncing branch (red) versus GR (blue).\textbf{Top right}: the solution for $h_f(\tau)$ at $k=100 H_0$ for the bouncing branch.}
			\label{fig:tbb100}
		\end{center}
\end{figure}

In the bouncing branch, the oscillations of $h_f$ are anti-damped for $\tau<\tau_b$ and damped for $\tau>\tau_b$ where $\tau_b$ is the bounce time corresponding to when $c(\tau)=0$. This happens at relatively late times, around $z_b \sim 0.6$. Until the bounce occurs, while $c<0$, $h_f$ experiences enormous growth, then starts to decay after $\tau_b$. Through the mixing term in \eqref{eq:tensorg}, this growth in $h_f$ translates into growth in the physical mode $h_g$, leading to oscillations that grow at late times. In FIG. \ref{fig:tbb100}, we show the evolution of the physical and dark sector tensors for equal amplitude initial conditions with $\tau_i = 10^{-6} H_0^{-1}$; the large deviation from GR at late times is clear. \\

The growth in $h_g(\tau)$ at late times is $k$-dependent. Empirically, we find a falloff proportional to $1/k^2$. This amplification is also  dependent on the initial time $\tau_i$, which physically relates to the reheating temperature $T_\text{reheat}=T_i$. 
This can be understood by examining the solution for $h_f$ within the radiation dominated era. On super-horizon scales there is an exact solution \cite{Lagos:2014aa} given by
\begin{equation} \label{eq:tau3}
	h_f = c_1 + c_2\tau^3
\end{equation}
so there is both a constant mode and a growing mode. Our choice of initial condition $\dot{h}_f=0$ selects the constant mode (ie. $c_2=0$), so naively one might think that $h_f$ should not grow at all in the radiation dominated era, regardless of the initial time $\tau_i$. However, the above solution \eqref{eq:tau3} is only an approximate solution on super-horizon scales, not valid for $k\neq 0$. For $k\neq 0$ we expect to depart from the the constant mode solution on a timescale of $\tau \sim 1/ck$, at which point the growing mode will completely dominate. The earlier the initial time is set, the more time $h_f$ has to grow, ultimately driving more growth in $h_g$. We find that the growth is inversely proportional to the initial time: $h_g(\tau_0) \propto 1/\tau_i \propto T_i$, valid for all initial times in the radiation dominated era. \\

In addition to this, the growth is also proportional to the initial condition for $h_f$. However, for a small enough value of $h_f(\tau_i)$, the solution for $h_g (\tau)$ does not scale. 
For our choice of initial time $\tau_i = 10^{-6}H_0^{-1}$ (corresponding to a reheat temperature of $T_i = 0.07$ GeV) we find that for $h_f(\tau_i)/h_g(\tau_i)<10^{-9}$, the solution for $h_g$ is indistinguishable from its solution with $h_f(\tau_i)=0$ which agrees very closely with the pure GR solution. To extrapolate this result to a more reasonable reheat temperature, say $T_i = 10^{10}$ GeV, we must consider that $h_f$ grows proportionally to $T_i$. We conclude that for  $T_i = 10^{10}$ GeV, we need a suppression of $h_f(\tau_i)/h_g(\tau_i)<10^{-20}$ to obtain solutions that agree with those of GR. At this threshold, scaling down $h_f(\tau_i)$ further will not result in any significant change. It is evident that to control the large growth in the bouncing branch, seeking gravitational waves that do not substantially deviate from those of GR, we require detailed knowledge about the mechanism by which they were produced. This will be explored further in Sec.~\ref{sec:initialconds}. \\  

\begin{figure}
		\begin{center}
			\subfigure{\includegraphics[width=7.5cm]{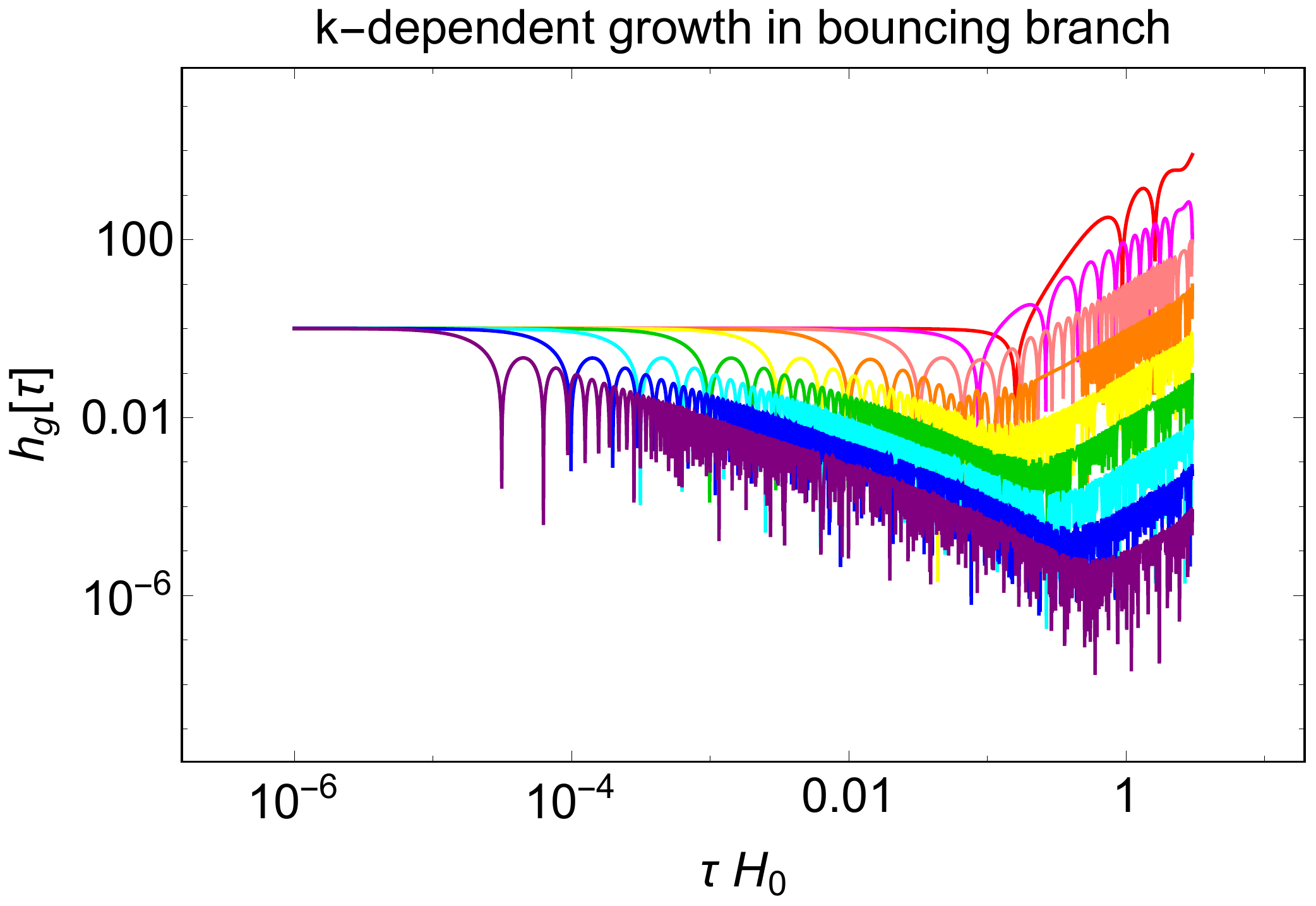}}
			\hspace{1cm}
			\subfigure{\includegraphics[width=7.5cm]{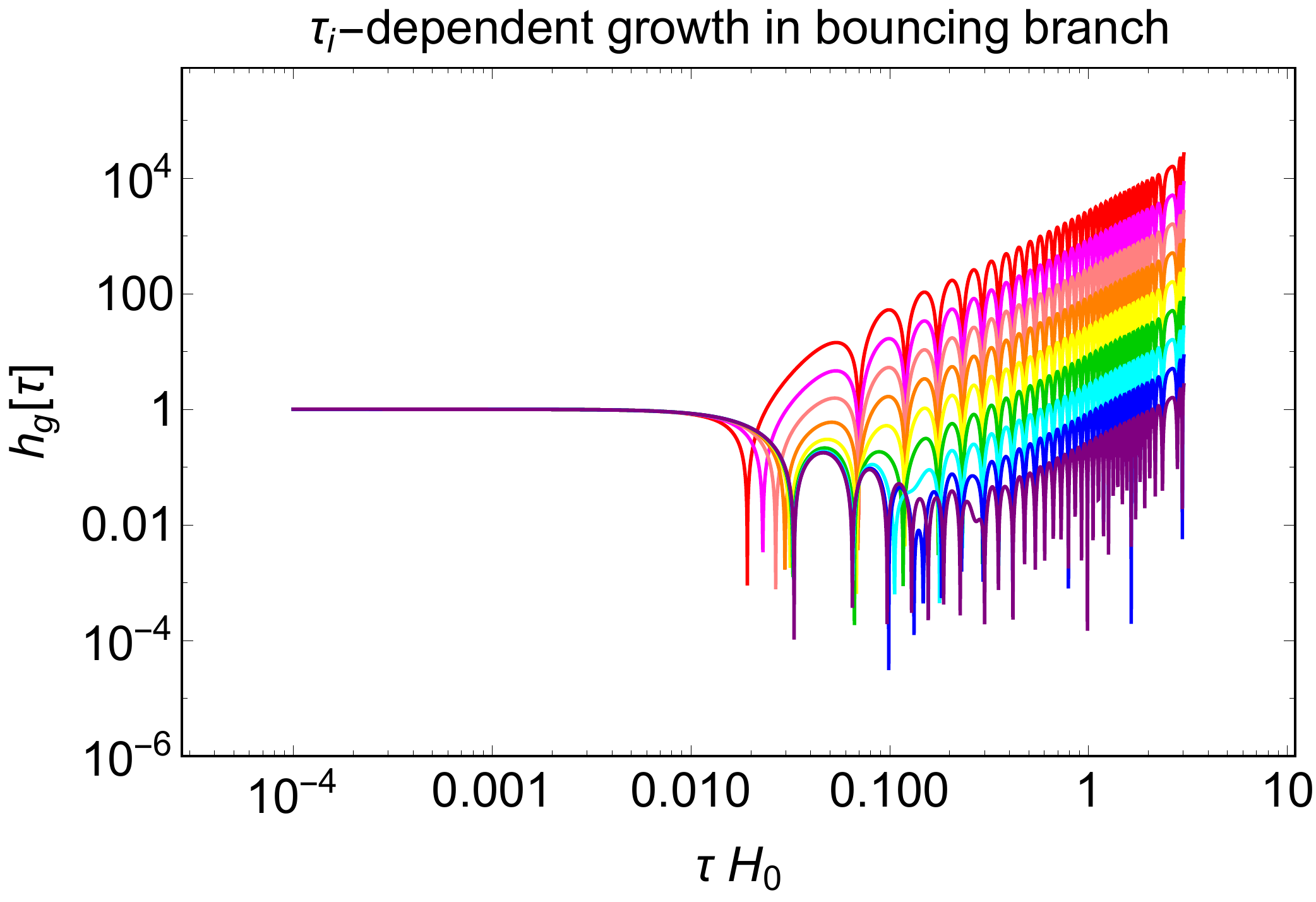}}
			\subfigure{\includegraphics[width=7.5cm]{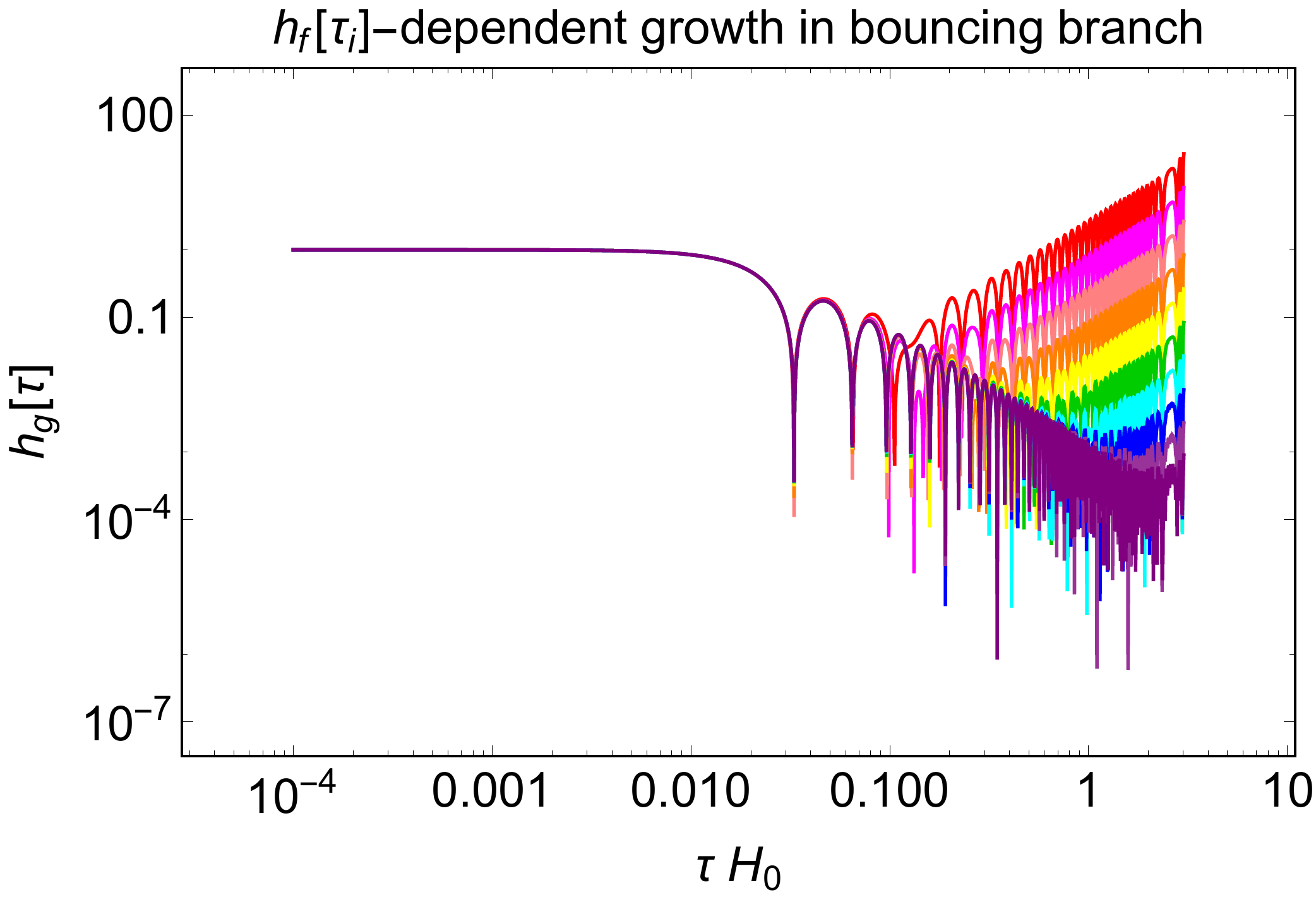}}
			\hspace{1cm}
			\subfigure{\includegraphics[width=7.5cm]{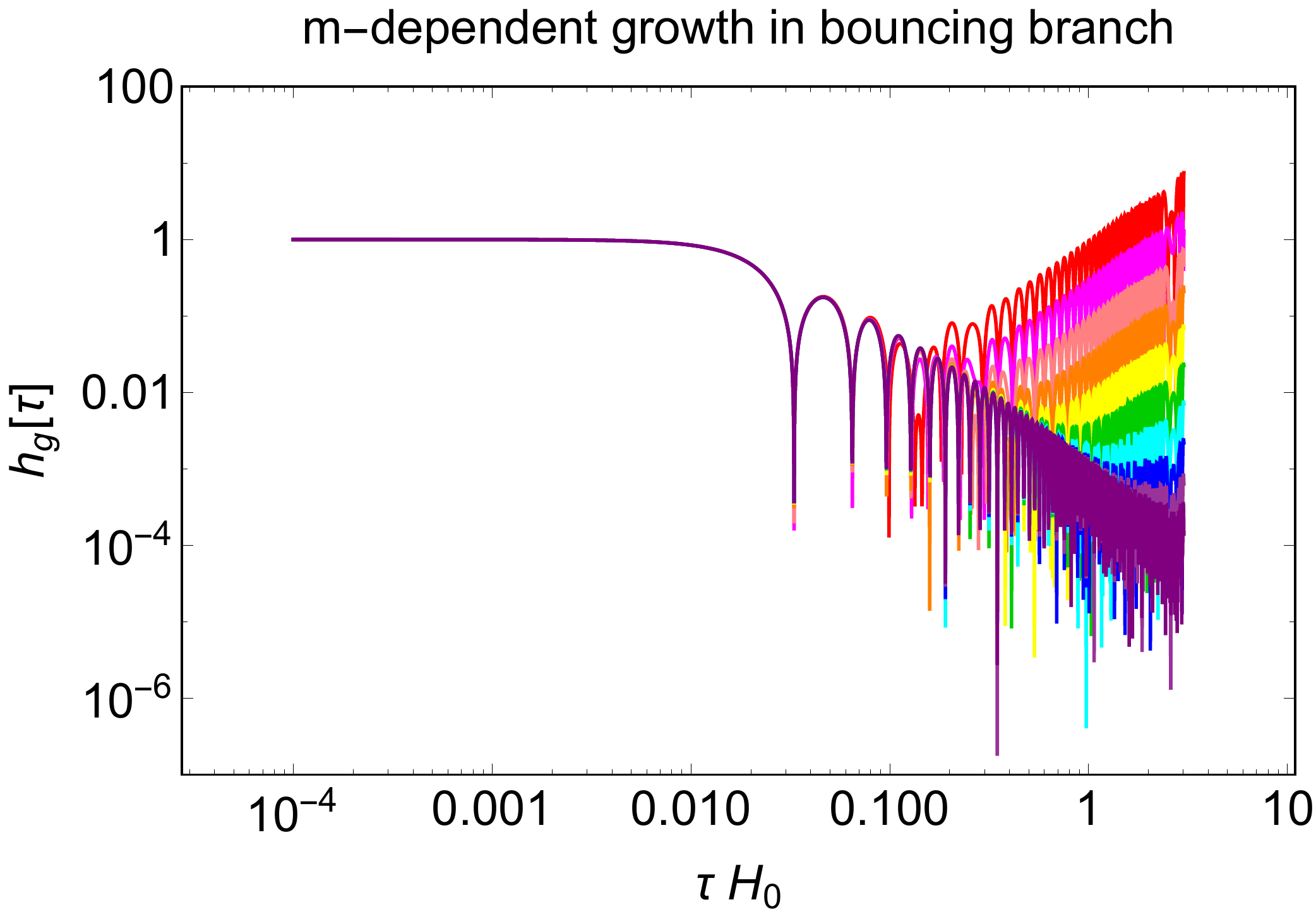}}
			\captionsetup{justification=raggedright,singlelinecheck=off,font=footnotesize}
			\caption{The growth in the physical sector is dependent on $k$, $\tau_i$, $h_f(\tau_i)$ and $m$ \eqref{eq:growth}. We vary one parameter individually per plot to show how the solutions change, fixing all but one of $\tau_i=10^{-6}H_0^{-1}$ ($T_i \sim 0.07$ GeV), $k=100 H_0$, $m=H_0$, and $h_{(g,f)}(\tau_i)=1$. \textbf{Top left}: The gravitational waves in the bouncing branch grow like $1/k^2$. Here we plot $k=10^{p/2}H_0$ with $p$ ranging from 2 to 10 from top to bottom. \textbf{Top right}: The tensor modes grow like $1/\tau_i$. Here we plot $\tau_i = 10^{-8+p/2}H_0^{-1}$ with $p$ ranging from $0$ to $8$ from top to bottom. \textbf{Bottom left}: The gravitational waves grow proportional to $h_f(\tau_i)$. Here we plot $h_f(\tau_i)=10^{-p/2}$ with $p$ ranging from 2 to 11 from top to bottom. \textbf{Bottom right}: The growth in the bouncing branch grow proportional to $m$ (requires re-inserting $\Lambda$ into the theory). Here we plot $m=10^{-p/2}H_0$ with $p$ ranging from 2 to 11 from top to bottom.}
			\label{fig:tbbgrowth}
		\end{center}
\end{figure}

From a purely phenomenological standpoint, one might also be interested in varying the graviton mass. This reduces the strength of the mixing term, which has the form $\propto m^2\beta^*_1a^2r(h_g-h_f)/H_0^2$ (reinserting factors of $m$ and $H_0$ from \eqref{eq:betas}). However, lowering $m$ means weakening the influence of the bigravity interaction term in \eqref{eq:action} to the point at which it can no longer yield a viable background cosmology, so a cosmological constant must be reintroduced. This undercuts the strongest theoretical motivation for bigravity, but mild variations could be of interest in constraining the parameters of the theory. We find that $h_g(\tau_0)/h_g(\tau_i) \propto m$. 
Just like for $h_f(\tau_i)$, there is also a critical value of $m$ for which this scaling relation no longer holds. 
Putting this all together, we find the following general relationship for the growth: 
\begin{equation}\label{eq:growth}
	 \frac{h_g(\tau_0)}{h_g(\tau_i)} \propto \frac{m h_f(\tau_i)}{\tau_i k^2} \propto \frac{m h_f(\tau_i)\mathcal{H}(\tau_i)}{k^2} \propto \frac{m h_f(\tau_i)T_i}{k^2},\ \ \ \ \ \text{for} \ \  \frac{h_f(\tau_i)}{h_g(\tau_i)}, \ m/H_0 \ >  A_{\rm crit}
\end{equation}
where $A_{\rm crit}$ is the critical value at which the solution no longer scales ($A_{\rm crit} = 10^{-20}$ for $ T_i=10^{10}$ GeV). Note that this scaling is different than that found in Cusin et. al., which is due to the difference in choices for initial data. These growth dependencies are displayed in FIG. \ref{fig:tbbgrowth} which shows the solutions for various $k$,  $\tau_i$, $h_f(\tau_i)$, and $m$. Notice that the variations in growth of $h_g$ are similar for $h_f(\tau_i)$ and $m$, which comes from the fact that both of these effects act to alter the coupling term in \eqref{eq:tensorg}. Although the effect on growth in the physical sector appears the same, altering $m$ introduces no change in growth in the dark sector. \\

As shown in Ref. \cite{Cusin:2014aa}, the generalized Higuchi bound in the tensor sector on an FRW background is satisfied for the expanding branch, but violated for the bouncing branch. This is essentially due to the behaviour of the lapse function, $c$, which takes on negative values in the bouncing branch, causing the kinetic term for the tensor perturbations to become negative. A violation of the Higuichi bound in an FRW background leads to a power law instability that is evident in our growing numerical solutions. This instability is not necessarily detrimental to the theory so long as these unstable modes are sufficiently suppressed, an important point which we investigate in section \ref{sec:initialconds}.

\section{CMB Tensor Power Spectrum} \label{sec:CMB}

In this section we use the solutions of \eqref{eq:tensorg} and \eqref{eq:tensorf} to compute the predicted bigravity CMB temperature tensor power spectrum (see \cite{Amendola:2015tua} and \cite{Enander:2015aa} for recent computations of CMB spectra in bigravity). We begin by using the physical tensor mode $h_{g}(k,\tau)$ to compute the $l$th photon moment due to tensor perturbations (assuming instantaneous recombination) using
\begin{equation} \label{eq:Theta}
	\Theta_l^T = -\frac{1}{2}\int_{\tau_*}^{\tau_0}{d\tau \  \dot{h}_g(k,\tau) j_l[k(\tau_0-\tau)]]}
\end{equation}
where $\tau_0$ is the conformal time today, $\tau_*$ is the conformal time at the time of last scattering, and $j_l$ is the spherical Bessel function. The tensor contribution to the temperature anisotropies is entirely due to the Integrated Sachs Wolfe effect. The evolution of the visible sector tensors can be substantially different than GR, leading to potentially visible differences in the spectrum of temperature anisotropies. For example, in the bouncing branch, from FIG.~\ref{fig:tbbgrowth} it can be seen that there is additional time dependence on super horizon scales, and late-time growth of tensor modes; both of these effects will alter the temperature anisotropies.\\

Once the photon moments are found, the angular power spectrum can be found from:
\begin{equation}\label{eq:ClT}
	C_l^T = \frac{(l-1)(l+1)(l+2)}{\pi}\int_0^\infty{dk \ \frac{1}{k} \ \biggl| \frac{\Theta_{(l-2)}^T}{(2l-1)(2l+1)} + 2\frac{\Theta_l^T}{(2l-1)(2l+3)} + \frac{\Theta_{(l+2)}^T}{(2l+1)(2l+3)} \biggl|^2}
\end{equation}
We solve for the photon moments and angular power spectra numerically. \\

For the expanding branch, the $C_l^T$'s are approximately the same as in GR, which is expected given the agreement of $h_g(\tau)$ with $h_\text{GR}(\tau)$. However, in the bouncing branch, there is a drastic difference in the power spectrum. \\

Starting with the measured value of the scalar quadrupole ${C_2^T}_\text{GR} \sim 1000 \mu K^2$, and saturating the bound on the tensor-to-scalar ratio of $r\equiv C_2^T/C_2^S<0.2$, the expected value of tensor quadrupole is ${C_2^T}_\text{GR} \sim 200\mu K^2$. In \eqref{eq:Theta} the integral runs from the time of last scattering to today, so the growth in $h_g(\tau)$ in the bouncing branch at late times causes a large increase in $\Theta_l^T$ and therefore in $C_l^T$. We find that if we set $h_g(\tau_i)=h_f(\tau_i)$ at $\tau_i=10^{-6}H_0^{-1}$ and $m=H_0$,  then the value of $C_2^T$ for the bouncing branch is $10^{11}$ times larger than for pure GR: ${C_2^T}_\text{BB} \sim 10^{11}{C_2^T}_\text{GR}$, demonstrating that the growth at late times dominates the signal. However, our initial time $\tau_i=10^{-6}H_0^{-1}$ corresponds to a reheat temperature of only $T_i=0.07$ GeV. Extrapolating to a more reasonable reheat temperature, say $T_i=10^{10}$ GeV, requires looking to equation \eqref{eq:growth} which shows that $h_g$ grows with $T_i$, and so we expect an even bigger enhancement of the quadrupole. \\

\begin{figure}[htb!]
		\begin{center}
			\subfigure{\includegraphics[width=5.5cm]{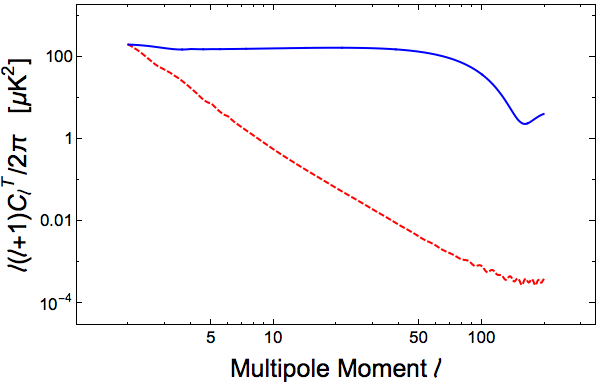}}
			\hspace{.5cm}
			\subfigure{\includegraphics[width=5.5cm]{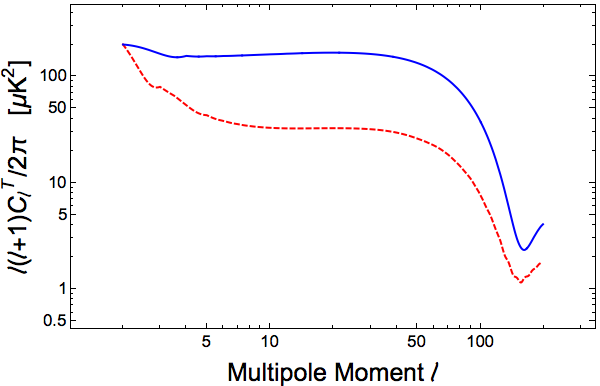}}
			\hspace{.5cm}
			\subfigure{\includegraphics[width=5.5cm]{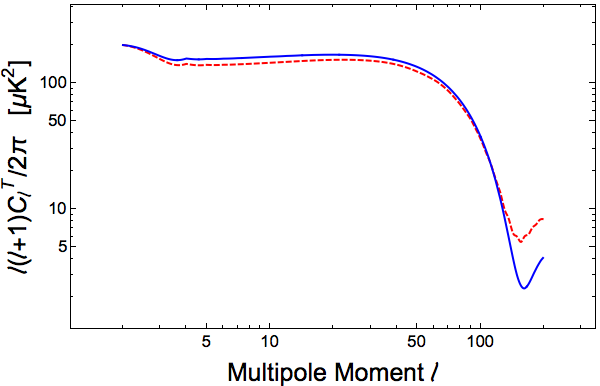}}
			\captionsetup{justification=raggedright,singlelinecheck=off,font=footnotesize}
			\caption{The CMB Tensor Power Spectrum in the bouncing branch (red, dashed) approaches the GR result (blue, solid) as the initial value of $h_f$ (or $m$ or $T_i$) is decreased from left to right: $h_f(\tau_i)/h_g(\tau_i) = 10^{-3},\ 10^{-5},\ 10^{-7}$ with $\tau_i=10^{-6}H_0^{-1}$. In each cases we have scaled the power spectrum in the bouncing branch down so that ${C_2^T}_\text{BB} \sim 200 \mu K^2$, which required dividing the spectrum by $\sim 10^6,\ 5.3,\ 1.1$ from left to right.}
			\label{fig:ClThfti}
		\end{center}
\end{figure}

To ensure that ${C_2^T}_\text{BB}={C_2^T}_\text{GR}\sim 200\mu K^2$ as phenomenologically required, we must divide the spectrum by a large factor, which is equivalent to a large suppression of the initial overall amplitude of tensor perturbations  $h_{(g,f)}(\tau_i)$. This can be accomplished by e.g. lowering the energy scale of inflation. In addition, by tuning the initial conditions, we can control the growth in $h_g$, and relieve the need for this large suppression factor. Referring to equation \eqref{eq:growth}, we see that an adjustment of $h_f(\tau_i)/h_g(\tau_i)$, $\tau_i$ (or $T_i$), or $m$ can lead to a power spectrum $C_l^T$ that is consistent with observations. For example, FIG. \ref{fig:ClThfti} shows how the bouncing branch power spectrum converges to the standard one as $h_f(\tau_i)/h_g(\tau_i)$ is decreased. Here, we have re-scaled the power spectrum to obtain $C_2^T/C_2^S \sim 0.2$. Note that these plots would look the same if instead, $m$ or $T_i$ was decreased by the same amount. In summary, to achieve a CMB Tensor Power Spectrum that resembles the result from GR, we require very tuned initial conditions, or tuned graviton mass. \\

\section{Present Day Stochastic Gravitational Wave Background} \label{sec:OmegaGW}

We now use the results from the previous section to see how the bigravity primordial gravitational waves contribute to the present day stochastic gravitational wave energy density . The observable quantity of interest is the gravitational wave energy density, defined as a function of frequency:
\begin{equation} 
	\Omega_\text{GW}^0(f)=\frac{1}{\rho_\text{crit}}\frac{d\rho_\text{GW}}{d\ln{f}}
\end{equation}
where the critical density is $\rho_\text{crit}=3M_\text{g}^2H^2(\tau)$. \\

Direct detection of relic gravitational waves is of considerable interest given the improving technology of ground and space based laser interferometers. Various experiments have already placed upper bounds on $\Omega_\text{GW}^0$, and proposed experiments will be able to reach much higher sensitivities. Therefore, one might ask if gravitational waves in bigravity would be more or less likely to detect, and if the current sensitives of LIGO or Pulsar Timing Arrays could constrain bigravity. LIGO has already made measurements between between $51<f<150$ Hz to constrain $\Omega^0_\text{GW}< 6.5 \times 10^{-5}$ at these frequencies and advanced LIGO is predicted to reach down to sensitivities of $\Omega^0_\text{GW}\sim 6.5 \times 10^{-9}$ in the coming years \cite{Abbott:2007aa}. In addition, Pulsar-timing experiments have placed an upper bound of $\Omega^0_\text{GW}< 1.6\times 10^{-9}$ at low frequencies $10^{-9}<f<10^{-8}$ Hz \cite{Jenet:2006aa} and will improve in the future. The first-generation space based laser interferometer, LISA, is expected to operate at sensitivities of $\Omega_\text{GW}^0 \sim 10^{-11}$ at frequencies $f\sim 10^{-3}$ Hz \cite{Ungarelli:2001aa}, while the second-generation space based interferometer, BBO, may be able to reach all the way down to $\Omega_\text{GW}^0 \sim 10^{-17}$ near frequencies $f\sim 0.3$ Hz \cite{Cutler:2006aa}. \\

In terms of the tensor modes (corresponding to the physical metric) the predicted stochastic background can be computed at any conformal time $\tau$ via the formula \cite{Boyle:2008aa}
\begin{equation} \label{eq:omegakt}
	\Omega_\text{GW}(k,\tau) = \frac{k^2|h_g(k,\tau)|^2+|\dot{h}_g(k,\tau)|^2}{12\pi^2\mathcal{H}^2(\tau)}
\end{equation}
given as a function of $k=2\pi f$. When written with superscript 0, it is understood to be evaluated today $\tau=\tau_0$. 
More generally however, $\Omega_\text{GW}^0(k)$ represents the present-day gravitational wave energy density on scales that re-entered the Hubble horizon during the radiation dominated era. 
\\

Over the range of frequencies of interest for the experiments above, the stochastic background due to primordial tensor modes in GR is essentially flat $\Omega_\text{GW}^0(k)\sim 10^{-15}$ \cite{Kuroyanagi:2014aa}. The precise profile depends on the assumed model of inflation that produced the modes, which for us is unimportant as we are just looking for a rough comparison to bigravity. \\

\begin{figure}[htb!]
		\begin{center}
			\subfigure{\includegraphics[width=7.5cm]{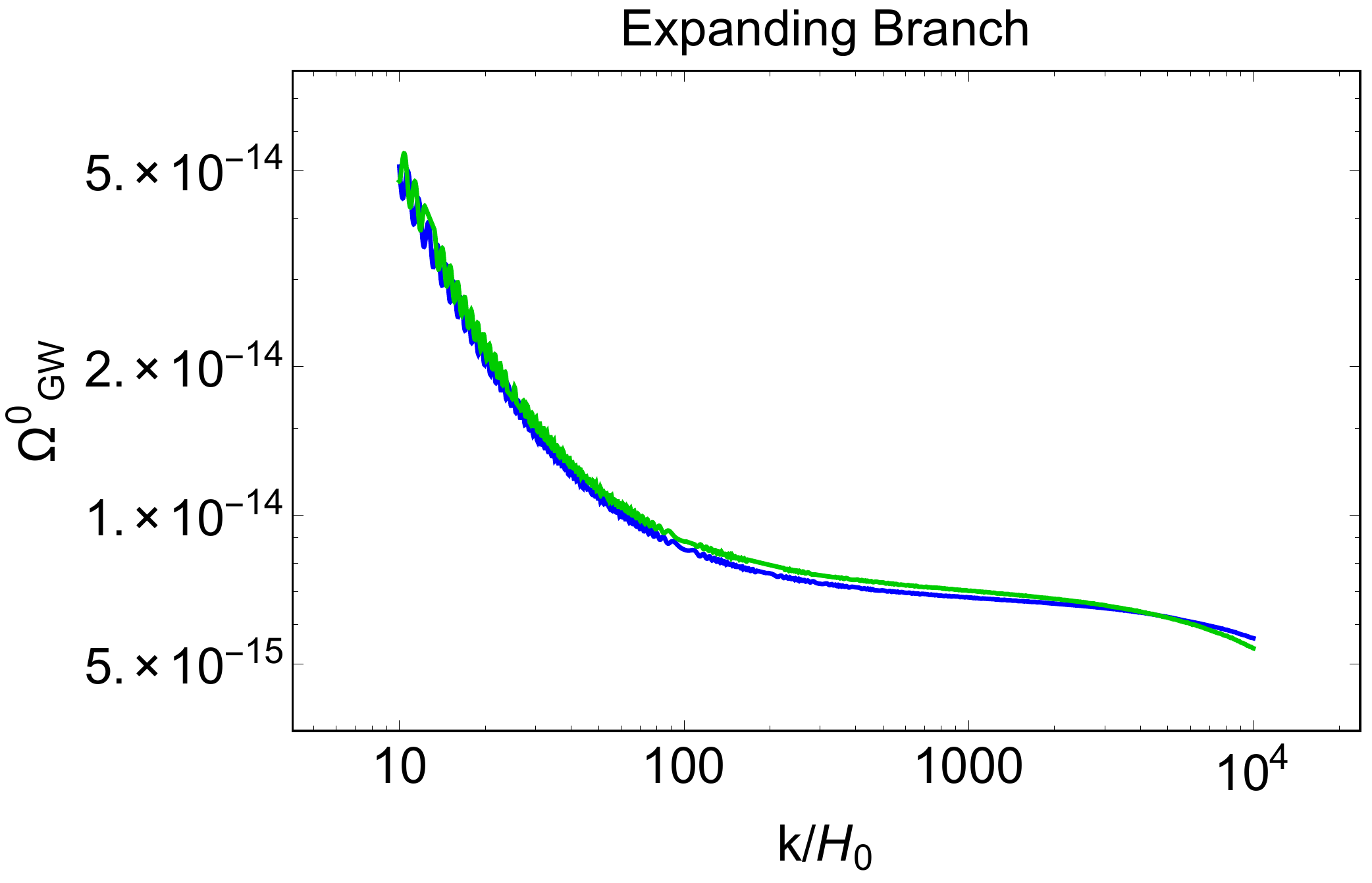}}
			\hspace{1cm}
			\subfigure{\includegraphics[width=7.5cm]{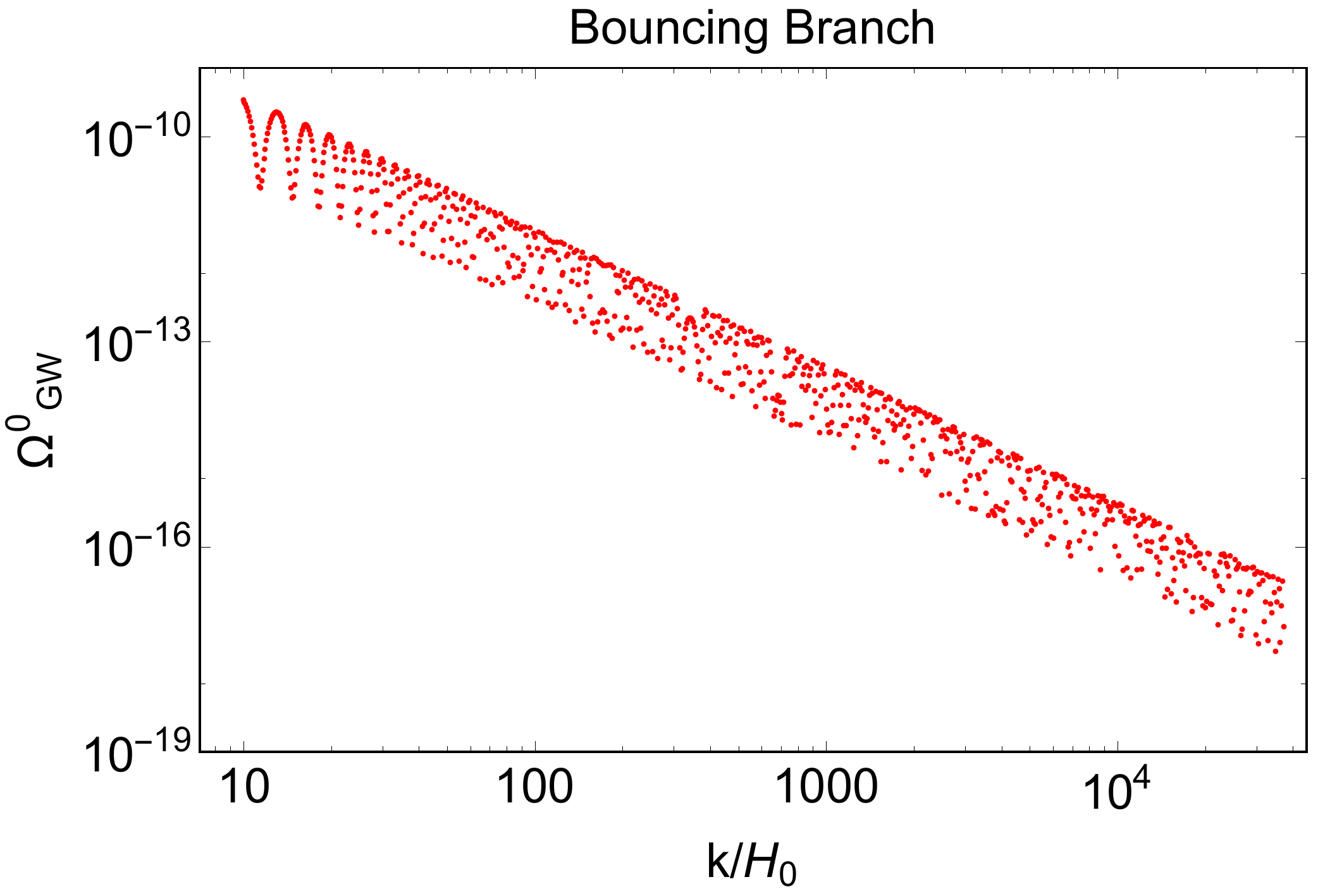}}
			\captionsetup{justification=raggedright,singlelinecheck=off,font=footnotesize}
			\caption{The present day stochastic gravitational wave background, given by \eqref{eq:omegakt} for bigravity as compared to GR. The expanding branch (green) agrees closely with GR (blue), while the bouncing branch (red) shows drastic differences. Note that we have scaled $\Omega_\text{GW}^0$ down by an appropriate factor so as to fix ${C_2^T}_\text{BB} \sim 200 \mu K^2$ (see FIG. \ref{fig:ClThfti}).}
			\label{fig:OGW}
		\end{center}
\end{figure}

Let us compare $\Omega_\text{GW}^0(k)$ in bigravity and GR. In the bouncing branch, we have observed that the late time growth of tensor modes falls of with the square of the frequency, and therefore, using \eqref{eq:omegakt}, so will $\Omega_\text{GW}^0(k)$. This makes $\Omega_\text{GW}^0(k)$ harder to detect in the bouncing branch compared to GR over the range of frequencies of interest for the experiments listed above, $10^{-9} < f < 10^3$ Hz (or $10^9 < k/H_0 <10^{21}$). In the expanding branch, the decay of the tensor modes closely matches with GR, so in this case we expect a result similar to the standard picture. See FIG. \ref{fig:OGW} for a plot of the results for $\Omega_\text{GW}^0$ over a range of frequencies from $k=10 H_0$ to $k=10^4H_0$.\\

\begin{figure}[htb!]
		\begin{center}
			\subfigure{\includegraphics[width=5.5cm]{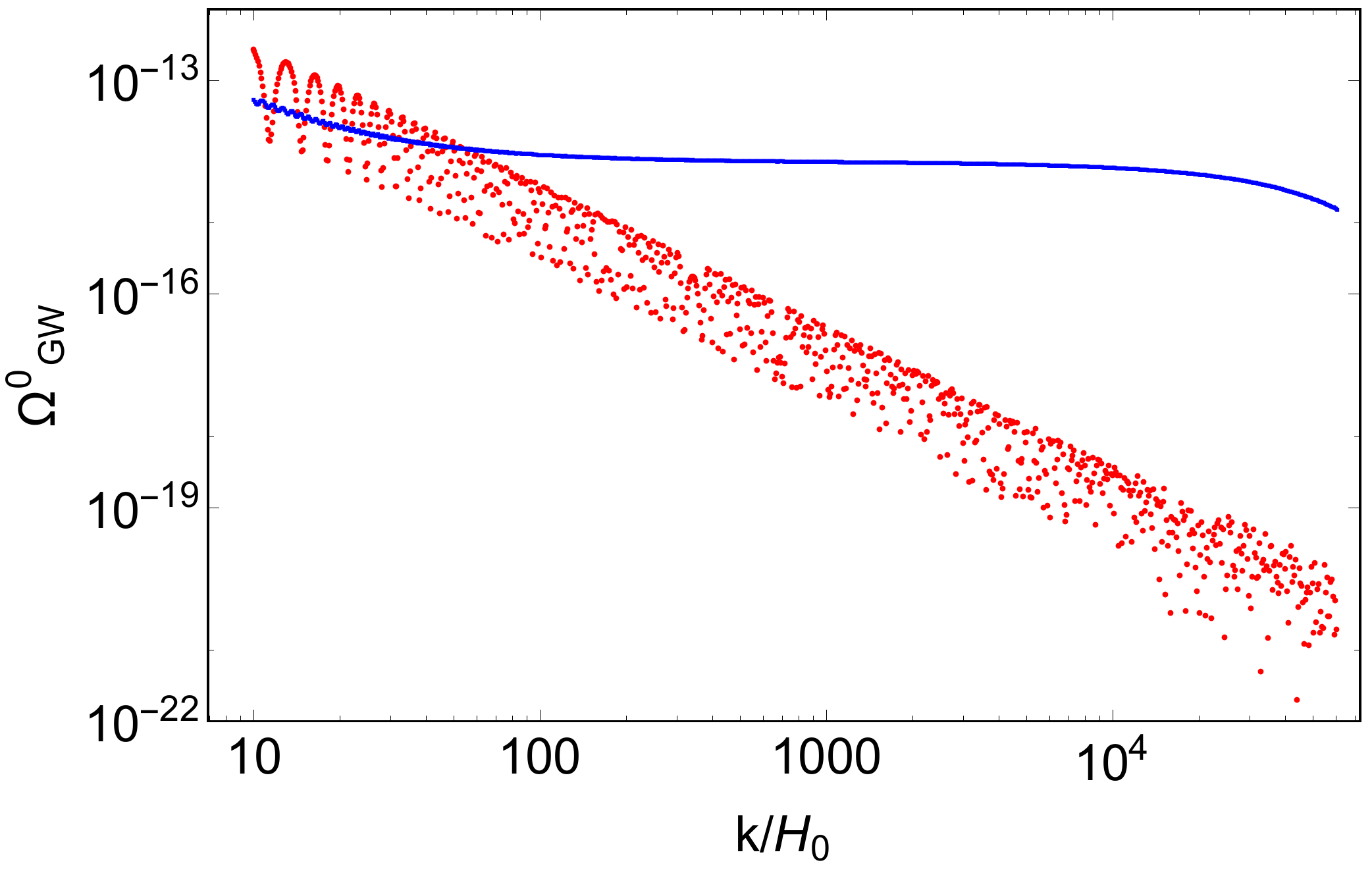}}
			\hspace{.5cm}
			\subfigure{\includegraphics[width=5.5cm]{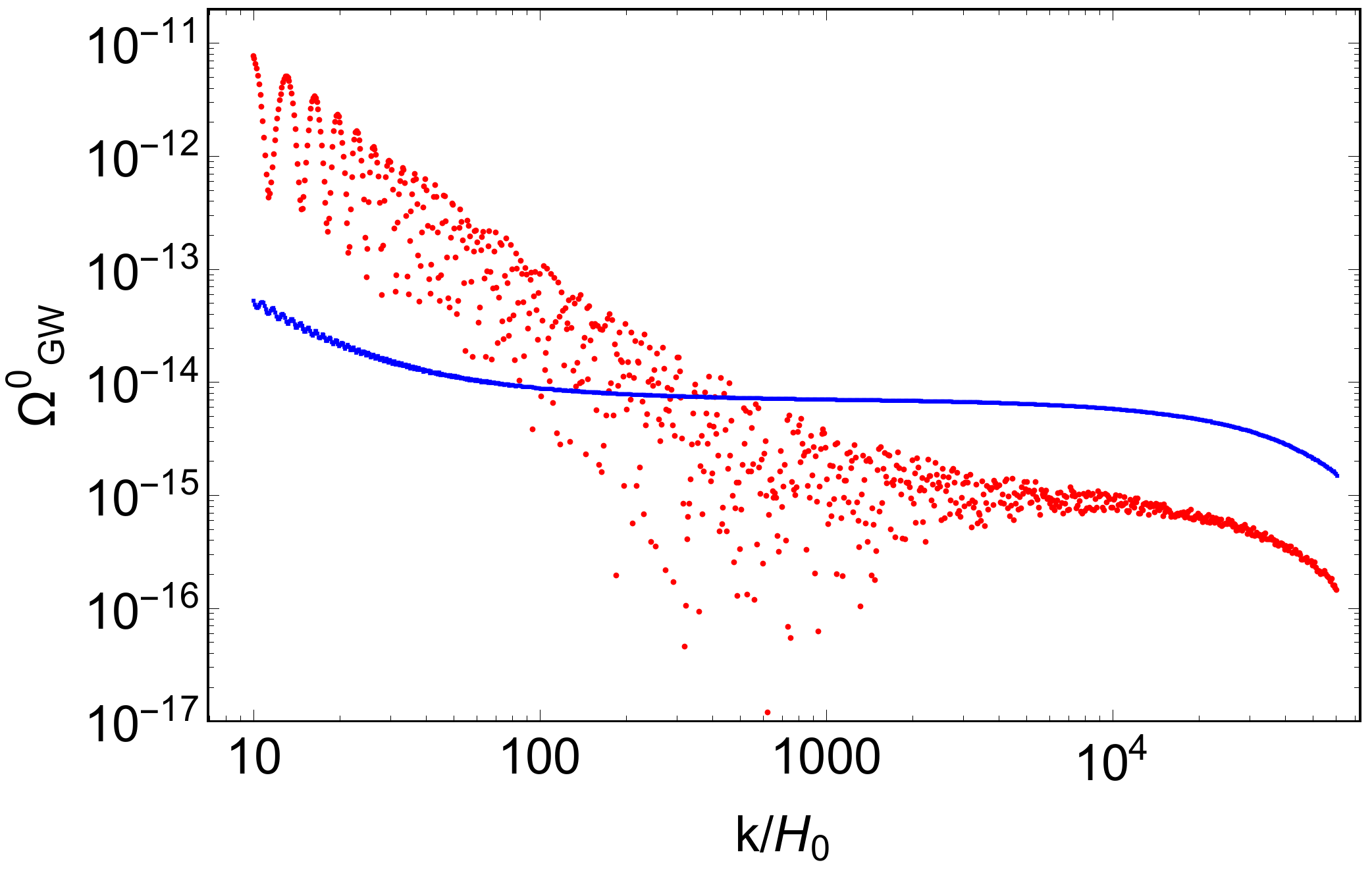}}
			\hspace{.5cm}
			\subfigure{\includegraphics[width=5.5cm]{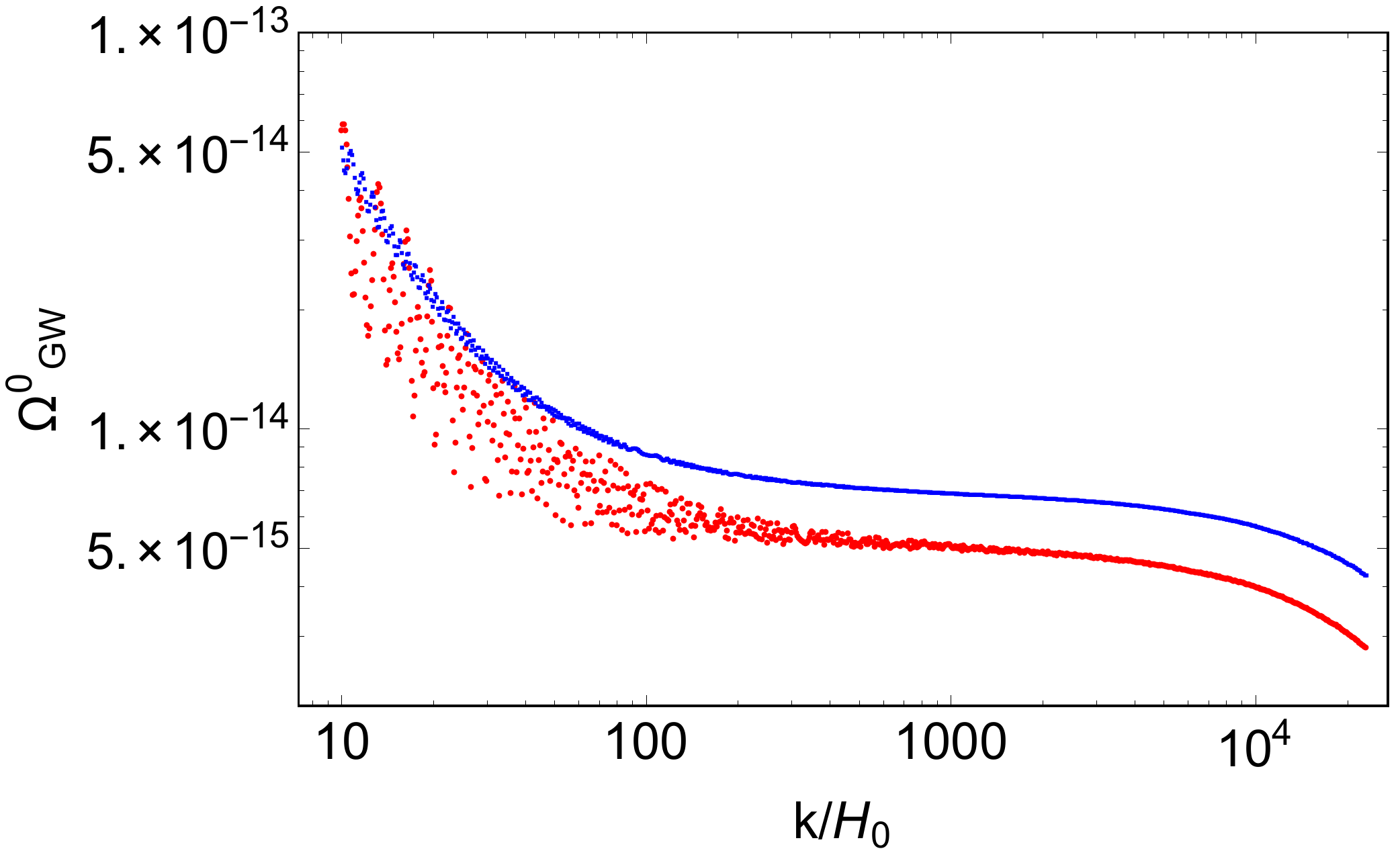}}
			\captionsetup{justification=raggedright,singlelinecheck=off,font=footnotesize}
			\caption{The present day stochastic gravitational wave background in the bouncing branch (red) approaches the GR result (blue) as the initial value of $h_f$ (or $m$ or $T_i$) is decreased from left to right: $h_f(\tau_i)/h_g(\tau_i) = 10^{-3},\ 10^{-5},\ 10^{-7}$ with $\tau_i=10^{-6}H_0^{-1}$}
			\label{fig:OGWhfti}
		\end{center}
\end{figure}

As in the previous section, we see that an adjustment of the initial condition for $h_f$ causes the result for the bouncing branch to converge to the solution in GR, as displayed in FIG. \ref{fig:OGWhfti}. This is equivalent to varying $m$ or $T_i$ by the same amount, as discussed previously. In this plot, we have re-scaled the power spectrum to yield a tensor contribution to the CMB temperature quadrupole of ${C_2^T}_\text{BB} \sim 200 \mu K^2$. \\

\section{Initial Conditions} \label{sec:initialconds}

We have seen that in the bouncing branch, the extreme growth in the dark sector causes amplification of the physical tensor mode, leading to large discrepancies with GR. This amplification causes alterations in physical observables, such as the CMB Power Spectrum and the present day stochastic gravitational wave background. However, if the tensor modes in the dark sector are sufficiently suppressed, then the physical tensor modes and their associated observables closely resemble those of GR. If some mechanism were to exist so that $h_f(\tau_i) \ll h_g(\tau_i)$ then this branch would have essentially the same gravitational wave spectrum as GR, and would be indistinguishable on an observational level. The need for this tuning has been observed in \cite{Lagos:2014aa,Cusin:2014aa}, and further in \cite{Amendola:2015tua}, where they estimate the required suppression to match the observed CMB power spectrum. It is therefore necessary to explore the initial conditions for the primordial tensor modes, assuming they were produced by inflation. \\

During inflation, the universe undergoes accelerated expansion in a quasi-de Sitter phase. For pure de Sitter, the bigravity background equations simplify as follows (see also Ref.~\cite{Sakakihara:2015aa}):
\begin{align}
	\rho & = \text{constant}=3 (H^I_g M_g)^2  \ \ \ \ \Rightarrow  \ \ \ \  \bar{\rho} = 3(H_g^I/H_0)^2\\
	r & = \text{constant} \\ \label{eq:lapseinf}
	c& = 1 \ \ \ \ \Rightarrow  \ \ \ \ \mathcal{H} = \mathcal{H}_g =\mathcal{H}_f 
\end{align}
where $H_g^I$ is the Hubble parameter during inflation. The second line follows from equation \eqref{eq:quarticr} and the third line follows from \eqref{eq:lapse}. Therefore, the dark universe is also undergoing de Sitter expansion. Restoring $m$ and $H_0$ in \eqref{eq:quarticr} using \eqref{eq:betas}, specializing to the bouncing branch parameters in which only $\beta_1^*$ and $\beta_4^*$ are nonzero, we get a polynomial equation for the value of the ratio of the scale factors during inflation, $r^I$: 
\begin{equation} \label{eq:rI}
-3(H_g^I)^2r^I + m^2\left[\beta_1^*+\beta_4^*(r^I)^3 -3\beta_1^*(r^I)^2\right]=0 .
\end{equation}
Since one typically takes the mass term $m$ in bigravity to be on the order of the Hubble constant $H_0$ which is much smaller than the Hubble parameter during inflation $H_g^I$, we can expand the solution for $r^I$ in powers of $H^I_g/m \gg 1$, yielding
\begin{eqnarray} \label{eq:ribb}
	r^I &=& \sqrt{\frac{3}{\beta_4^*}}\frac{H^I_g}{m} + \frac{3\beta_1^*}{2\beta_4^*} + \mathcal{O}\left( \frac{m}{H^I_g}\right)^2 \\
	&\sim& \frac{H^I_g}{H_0} 
\end{eqnarray}
where we have used $m = H_0 \ \Rightarrow \ \beta^*_4=\beta_4=0.94$. Assuming high scale inflation, the maximum allowed Hubble during inflation is $H^I_g \sim 10^{15}$ GeV, and $H_0 \sim 10^{-33}$ eV, from which we can estimate $r^I \sim 10^{57}$. For very low scale inflation, say at the TeV scale, we can estimate $r^I \sim 10^{29}$. \\

After inflation, there must be a transition to a radiation dominated phase. During the inflationary de Sitter phase, the scale factor for the dark sector metric is increasing, $\dot{b}>0$, but in the radiation dominated era in the bouncing branch, the scale factor is decreasing, $\dot{b}<0$. It is evident that the dark sector must undergo another ``bounce" transition after inflation from expansion to contraction in order to achieve the necessary behaviour in the early radiation era. In the next section we see that this bounce is indeed achieved in a simple inflationary model. As argued for the late-time bounce in this branch, it is likely that the associated divergence in the curvature is a mathematical feature rather than a physical problem.\\

A note on inflation in the expanding branch is now in order. In this branch, we set $\beta_4^*=0$. Solving for $r^I$ in \eqref{eq:rI} with only $\beta_1^*\neq 0$, we obtain
\begin{eqnarray}
	r^I &=& \frac{-3{H_g^I}^2+\sqrt{9{H_g^I}^2+12m^4{\beta^*_1}^2}}{6m^2\beta_1^*} \\
	     &=& \frac{\beta_1^*m^2}{3{H_g^I}^2} + \mathcal{O}\left( \frac{m}{H_g^I} \right)^3 \\
	     &\sim& \left( \frac{H_0}{H_g^I} \right)^2
\end{eqnarray}
Therefore, in the expanding branch we obtain a very small value of $r^I$. \\

To find the power spectrum of primordial tensors, we expand the action Eq.~\ref{eq:action2} to quadratic order in transverse traceless perturbations of the $f$ and $g$ metrics. During inflation, the interaction terms are unimportant due to the large hierarchy between $m$ and $H_g^I$. Defining the canonically normalized fields:
\begin{equation}\label{eq:mode_fns1}
v_g = \frac{M_g a}{2} h_g, \ \ \ v_f = \frac{M_g b}{2} h_f,
\end{equation}
and using the fact that
\begin{equation}
\frac{\ddot{a}}{a} = \frac{\ddot{b}}{b} = \frac{2}{\tau^2} 
\end{equation}
in de Sitter, the quadratic action for tensors during an inflationary epoch in bigravity is given by:
\begin{equation} \label{eq:actionquad}
	S =  \sum_{+, \times} \frac{1}{2} \int d\tau \ d^3 k \left[ \dot{v}_{g,f}^2 - \left( k^2 - \frac{2}{\tau^2} \right) v_{g,f}^2  \right]
\end{equation}
Imposing Bunch Davies initial conditions, the mode functions each obey:
\begin{equation}\label{eq:mode_fns2}
v_{g,f} = \frac{1}{k^{3/2} \tau} \left( 1 - i k \tau \right) e^{i k \tau}
\end{equation}

Summing over the two polarization states, the power spectrum for $h_{f}$ and $h_g$ are given by:
\begin{equation}
	{\Delta_T^{(g)}}^2 = \left . \frac{2 k^3}{\pi^2 M_g^2} \frac{|v_{g}|^2}{a^2} \right|_{k = \hat{\mathcal{H}}}, \ \ \  {\Delta_T^{(f)}}^2 = \left . \frac{2 k^3}{\pi^2 M_g^2} \frac{|v_{f}|^2}{b^2} \right|_{k = \hat{\mathcal{H}}}
\end{equation}
Substituting with Eqn.~\ref{eq:mode_fns2} and using $H_{g}^{I} = (a\tau)^{-1}$ for a de Sitter phase, we obtain:
\begin{equation} \label{eq:rIsuppression}
{\Delta_T^{(g)}}^2 = \left .\frac{2 {H_{g}^{I}}^2}{\pi^2 M_g^2}  \right|_{k = \hat{\mathcal{H}}} , \ \ {\Delta_T^{(f)}}^2 = \frac{{\Delta_T^{(g)}}^2}{{r^I}^2}  
\end{equation}
Modes are populated for both the visible and dark sector tensors on all super horizon scales. The appearance of $r^I$ is a consequence of the fact that the relative size of $a$ and $b$ is physical, and cannot be removed by a change of coordinates or field redefinition. \\

In the bouncing branch $r^I \gg 1$, leading to a drastic suppression in the initial amplitude for $h_f$ roughly given by $h_f / h_g = 1 / r^I \sim H_0 / H^I_g$. This suppression is more than sufficient to bring the amplitude of dark sector tensor fluctuations below the threshold where they alter the propagation of visible sector tensors. With inflationary initial conditions, we therefore conclude that there would be no visible deviation from GR in the tensor contribution to the CMB temperature anisotropies or the late time stochastic distribution of gravitational waves. This is true for both high scale or low scale inflation. \\

In contrast, the expanding branch gives $h_f / h_g \sim (H^I_g/H_0)^2$, which for any reasonable choice of the inflationary scale is far beyond the perturbative regime. Therefore, we cannot make sense of inflation in the expanding branch with $\beta_4=0$. If instead we considered a non-minimal expanding branch with $\beta_4 \neq 0$, then the result would be the same as the bouncing branch. \\

Finally, let us comment on the time dependence of the dark sector tensor modes. This is of interest because as illustrated in \cite{Lagos:2014aa,Cusin:2014aa}, the dark sector tensor mode has a growing mode on super horizon scales proportional to $\tau^3$ during radiation domination. We can estimate the relative amplitude of the growing and constant modes of \eqref{eq:tau3} at the beginning of radiation domination as $|\tau  \dot{h}_f (\tau_i) / h_f (\tau_i)|$. Using \eqref{eq:mode_fns1} and \eqref{eq:mode_fns2}, one obtains
\begin{equation}
|h_f(\tau_i)| = \frac{2}{M_g}\frac{(1+k^2\tau_i^2)^{1/2}}{k^{3/2}\tau_i b(\tau_i)} = \frac{2}{M_g}\frac{H_g^I(1+k^2\tau_i^2)^{1/2}}{k^{3/2} r^I} \hspace{1cm} \Rightarrow \hspace{1cm} |\dot{h}_f(\tau_i)| =  \frac{2}{M_g} \frac{H_g^I k^2\tau_i}{2k^{3/2}r^I(1+k^2\tau_i^2)^{1/2}},
\end{equation}
where we used that $b(\tau_i)=r^Ia(\tau_i)=r^I/(H_g^I\tau_i)$. In the limit of small $\tau_i$, the ratio between the growing and constant mode is therefore
\begin{equation}
\biggl| \frac{\tau_i \dot{h}_f (\tau_i)}{h_f (\tau_i)} \biggl|= \frac{k^2\tau_i^2}{2}=\frac{k^2 \tau_i}{2 a (\tau_i) H_g^I} .
\end{equation}
Using $a \propto \tau$ during radiation domination, for horizon-scale wave numbers $k\sim H_0$, in the bouncing branch we can estimate this ratio as 
\begin{equation}\label{eq:velocity_condition}
\biggl| \frac{\tau_i \dot{h}_f (\tau_i)}{h_f (\tau_i)} \biggl|\sim \frac{H_0^2 \tau_i}{(\tau_i/\tau_0)H_g^I} \sim \frac{H_0}{H_g^I}\sim 10^{-57}
\end{equation}
For any reasonable choice of reheat temperature, this will be smaller than the growth factor for the growing mode during radiation domination, $\sim (\tau_{\rm eq} / \tau_i)^3 = (T_{\rm reh} / T_{\rm eq})^3 \sim 10^{30}$, where the subscript ``eq" refers to matter radiation equality. We conclude that the growing mode remains significantly suppressed, and therefore the appropriate initial conditions are $\dot{h}_{g,f} = 0$. \\

Before moving on to a specific inflationary model in bigravity, let us comment on the validity of these calculations. Massive bigravity is an effective field theory with a strong coupling scale $\Lambda_3 = (m^2M_g)^{1/3}$. For $m\sim H_0$, this scale is rather low:  $\Lambda_3 \sim 10^{-22}$ GeV. This is much smaller than any reasonable inflationary scale. Therefore, we should worry about the validity of the effective field theory during inflation.  While the scalar perturbations become strongly coupled, the potential term for the transverse traceless modes becomes irrelevant during inflation. However, we should be concerned about interactions between the scalar and transverse traceless perturbations which may contribute to the evolution of the tensors. A possible resolution to describing inflation in bigravity lies in the Vainshtein screening mechanism \cite{Vainshtein:1972sx}. If screening is efficient in the high density environment of inflation, the scale strong coupling scale gets redressed to a far higher scale $\Lambda_3^* \gg \Lambda_3$, and the structure of the potential is not affected (see, for instance \cite{de-Rham:2013ab,Heisenberg:2014aa}). Under this assumption of screening, calculations at the inflationary scale might still be valid in the effective field theory description of bigravity. A rigorous proof of this argument is still required, which we leave to future work.

\subsection{An inflationary model in bigravity}

In this section, we examine bigravity for the $m^2\phi^2$ inflationary model. We want to determine the behaviour of $r$ (and thus $b$) during inflation in the bouncing branch with only $\beta_{1,4} \neq 0$. The inflaton field has potential energy $V(\phi) = \frac{1}{2}m_\phi^2\phi^2$ and energy density $\rho_\phi = \frac{1}{2}(\partial_t\phi)^2 +V(\phi)$. For this calculation we find it convenient to use the following set of dimensionless variables:
\begin{eqnarray}
	\tilde{H} &=& \frac{H}{H_g^I} \hspace{1cm} \tilde{t} = t H_g^I \hspace{1cm} \tilde{\Gamma}_\phi = \frac{\Gamma_\phi}{H_g^I}   \hspace{1cm} \tilde{V} = \frac{V}{M_g^2(H_g^I)^2}\\
	\tilde{\phi} &=& \frac{\phi}{M_g} \hspace{1cm} \tilde{\beta}_n = \frac{m^2}{(H_g^I)^2}\beta_n^* = \frac{H_0^2}{(H_g^I)^2}\beta_n \hspace{.6cm} \tilde{\rho}_r = \frac{\rho_r}{M_g^2(H_g^I)^2}
\end{eqnarray}
where $\Gamma_\phi$ is the decay rate of the inflaton, $t$ is proper time, and we take $(H_g^I)^2 = V(\phi_0)/3M_g^2$ in terms of the value of $\phi$ at the start of inflation, implying that $\tilde{V} = 3\tilde{\phi}^2/\tilde{\phi}_0^2$. \\

Including an explicit decay of the inflaton into radiation, the equation of motion is 
\begin{equation} \label{eq:eomphi}
	\tilde{\phi}'' + 3\tilde{H}\tilde{\phi}'+\tilde{\Gamma_\phi}\tilde{\phi}'+6\frac{\tilde{\phi}}{\tilde{\phi_0}^2}=0
\end{equation}
where a prime denotes a derivative with respect to dimensionless proper time $\tilde{t}$. The Friedmann equation \eqref{eq:eomg1} for $a(\tilde{t})$ becomes
\begin{equation} \label{eq:eoma}
	3\tilde{H}^2 = \tilde{\rho}_\phi + \tilde{\rho}_r = \frac{1}{2}(\tilde{\phi}')^2+3\frac{\tilde{\phi}^2}{\tilde{\phi}_0^2} + \tilde{\rho}_r .
\end{equation}
Notice that we have neglected the contribution of $\tilde{\rho}_m$ and $\tilde{\rho}_\text{mg}$ in the early universe since these will be highly suppressed compared to the inflaton or radiation energy density. The radiation energy density $\tilde{\rho}_r$ satisfies a modified conservation equation
\begin{equation} \label{eq:eomrhor}
\tilde{\rho}_r'+4\tilde{H}\tilde{\rho}_r = \tilde{\Gamma}_\phi\tilde{\rho}_\phi
\end{equation}
We can now solve \eqref{eq:eomphi}, \eqref{eq:eoma}, and \eqref{eq:eomrhor} for the functions $\tilde{\phi}(\tilde{t}),\ a(\tilde{t}),\ \tilde{\rho_r}(\tilde{t})$. The last ingredient will be to solve for $r(\tilde{t})$, for which we use \eqref{eq:quarticr}, which simplifies in the bouncing branch with $\beta_{1,4} \neq 0$ to
\begin{eqnarray} 
	0 &=& r \tilde{\rho} - \tilde{\beta}_1 - \tilde{\beta}_4 r^3 + 3 \tilde{\beta}_1 r^2 \\ \label{eq:rapprox}
	r &=& \frac{H_g^I}{H_0}\sqrt{\frac{\tilde{\rho}}{\beta_4}} \ \ \ \ \text{for  large $r$}
\end{eqnarray}

\begin{figure}[htb!]
		\begin{center}
			\subfigure{\includegraphics[width=7.5cm]{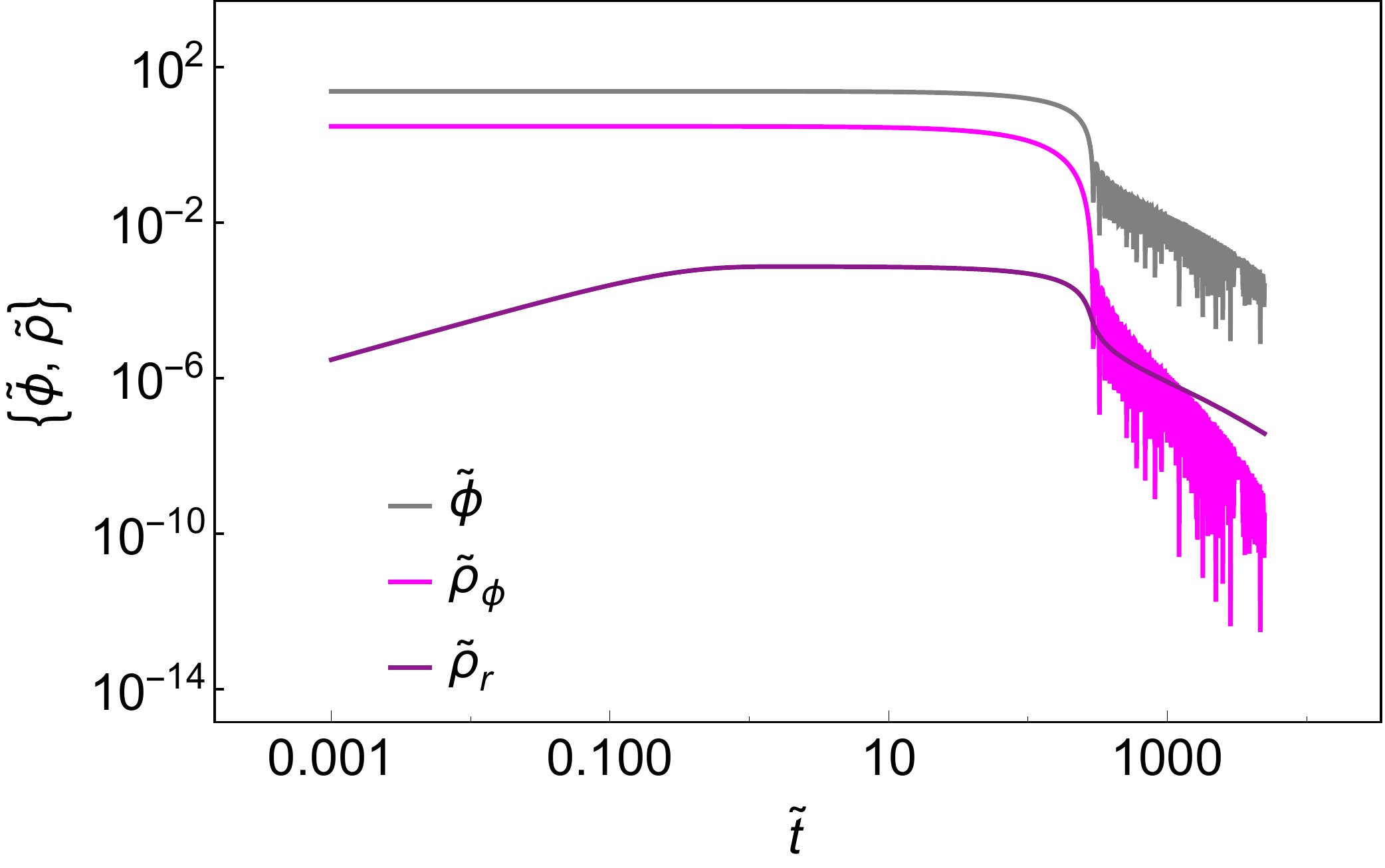}}
			\hspace{.5cm}
			\subfigure{\includegraphics[width=7.5cm]{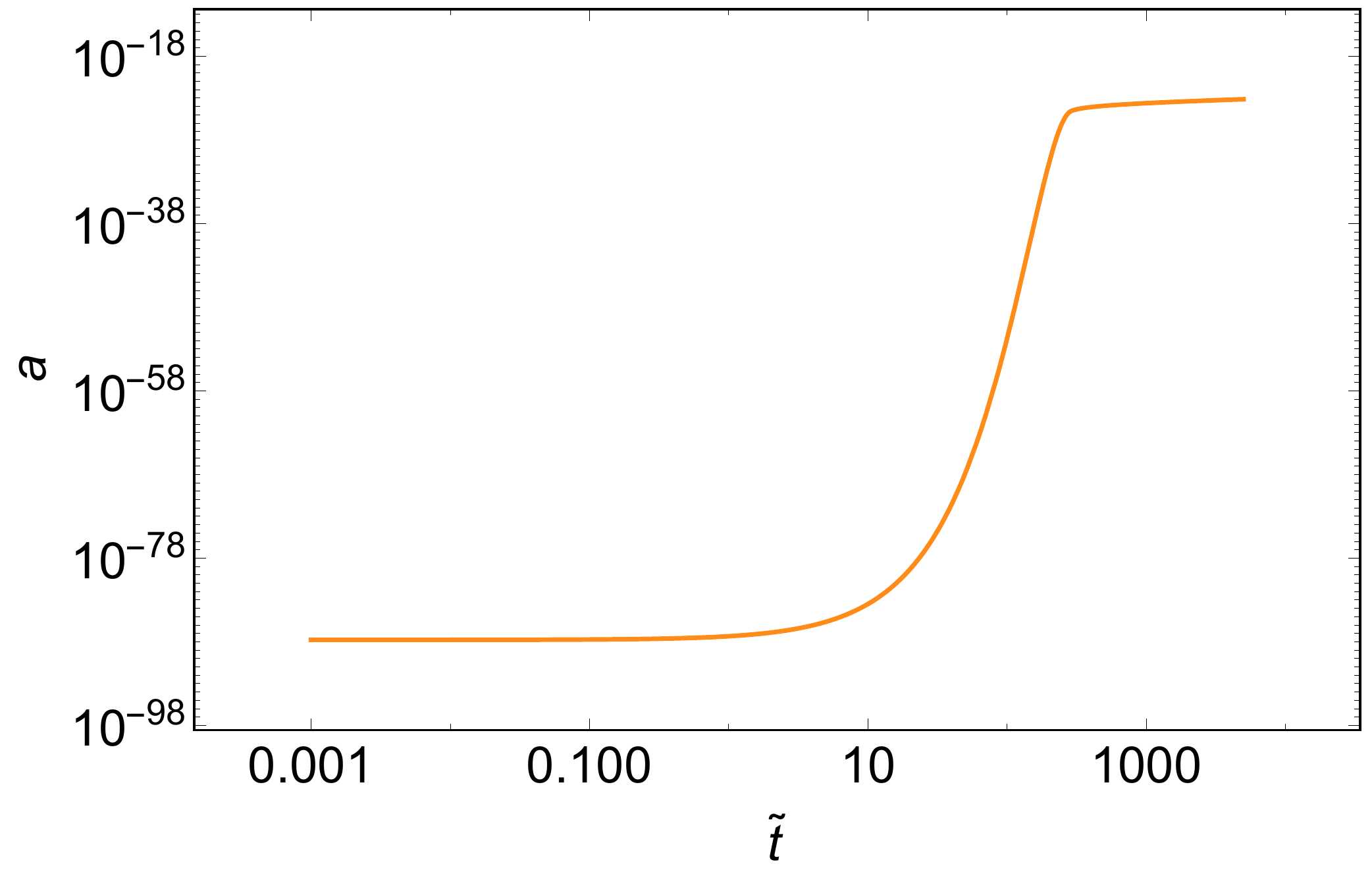}}\\
			\subfigure{\includegraphics[width=7.5cm]{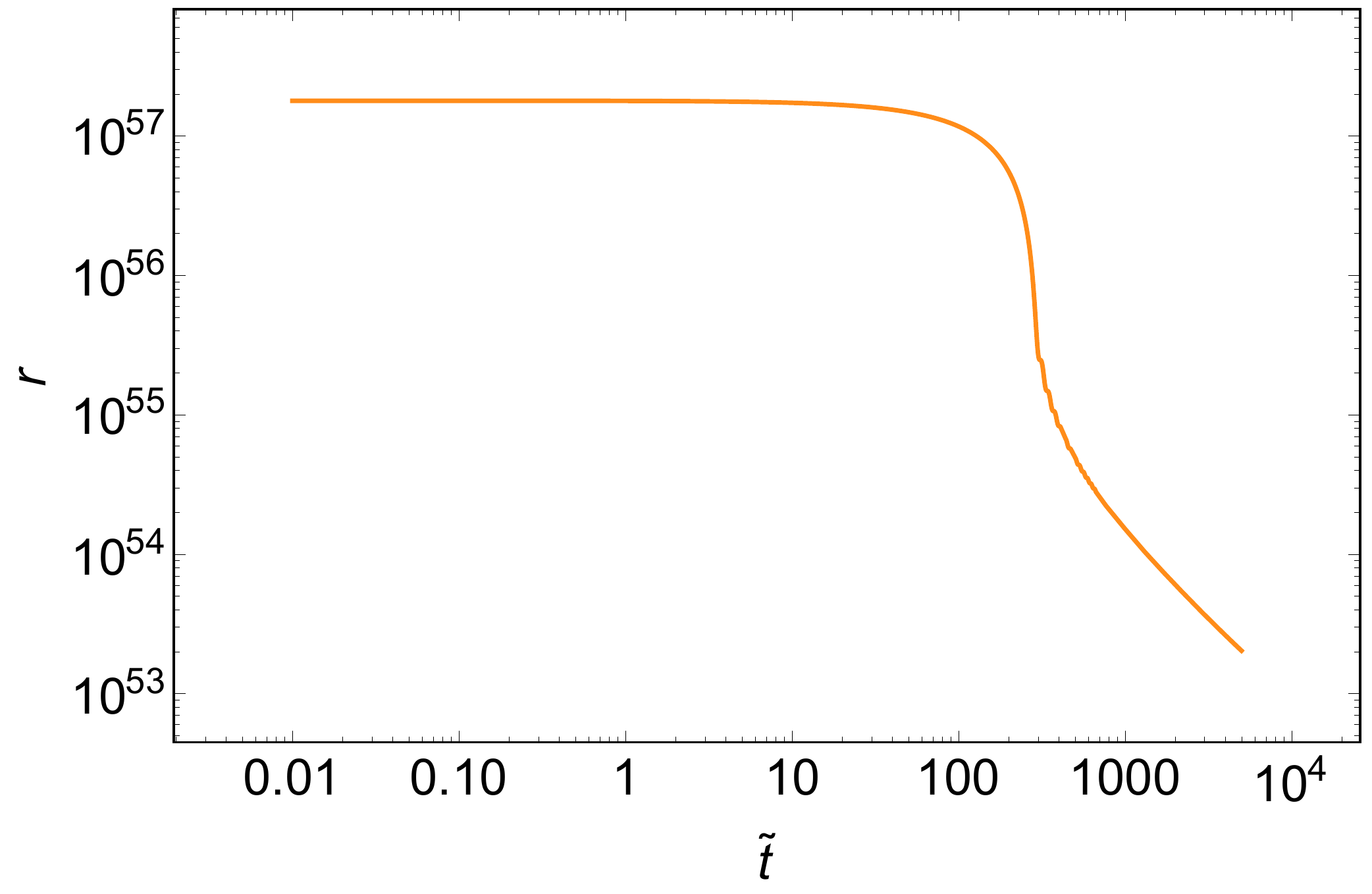}}
			\hspace{.5cm}
			\subfigure{\includegraphics[width=7.5cm]{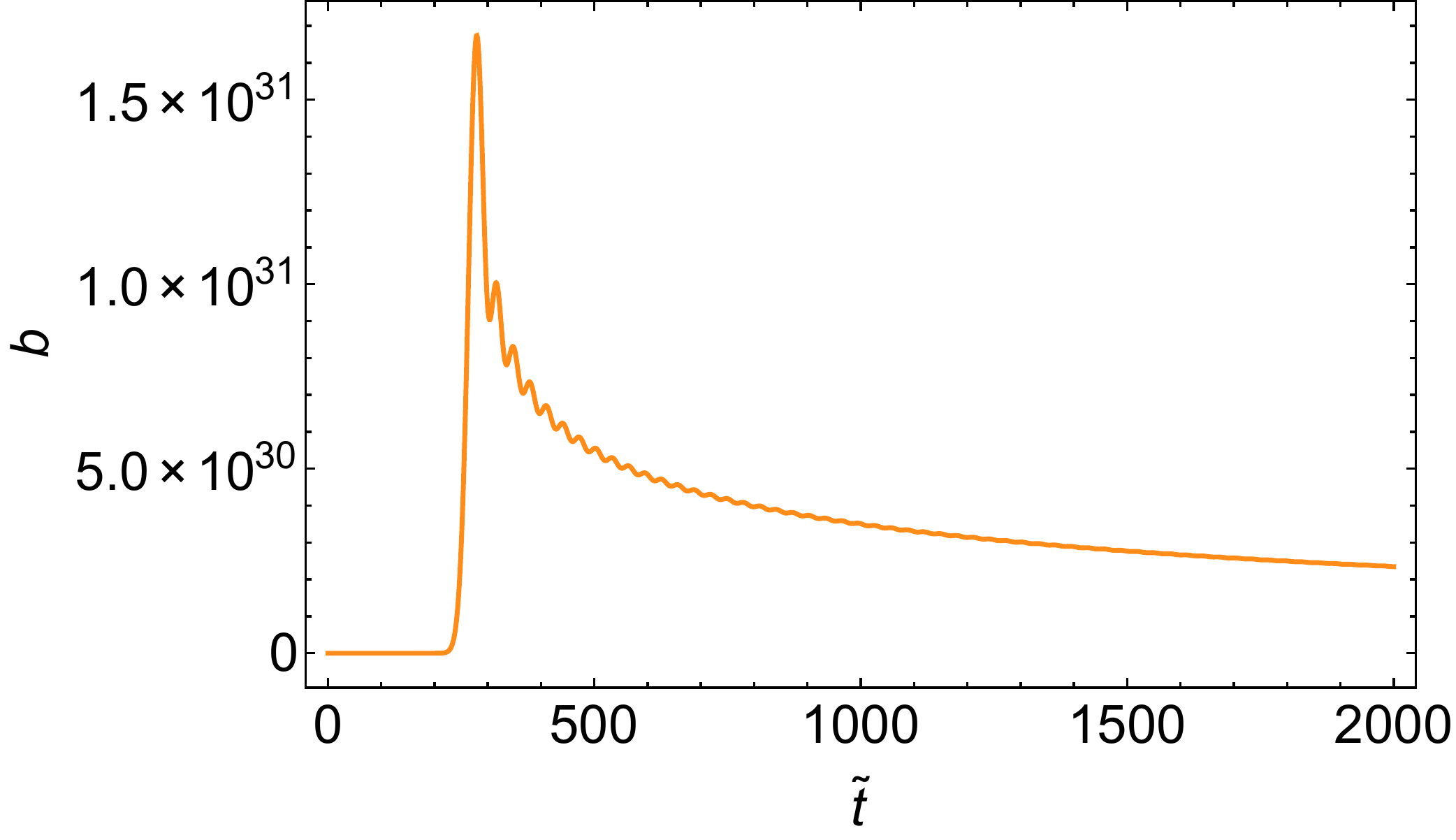}}
			\captionsetup{justification=raggedright,singlelinecheck=off,font=footnotesize}
			\caption{\textbf{Top left}: The solutions of \eqref{eq:eomphi}, \eqref{eq:eoma}, and \eqref{eq:eomrhor} for $\tilde{\phi},\ \tilde{\rho}_\phi$, and $\tilde{\rho}_r$ vs. $\tilde{t}$ for the $m^2\phi^2$ model of inflation. We set $\Gamma_\phi=10^{-3}$,  $\tilde{\phi}(0)=\tilde{\phi}_0 = 24$, $\tilde{\phi}'(0)=0$, and $\tilde{\rho}_r(0)=0$. \textbf{Top right}: The scale factor for $g_{\mu\nu}$. \textbf{Bottom left}: The ratio $r$ found via \eqref{eq:rapprox}. \textbf{Bottom right}: The scale factor for $f_{\mu\nu}$, $b = r a$. }
			\label{fig:inflation}
		\end{center}
\end{figure}

The results of the calculation are shown in FIG. \ref{fig:inflation}. We can see the transition from $b'>0$ during inflation to $b'<0$ after inflation is achieved in the bottom right plot. Notice that after $b$ hits its first peak, it oscillates as it decreases, indicating brief periods of expansion and contraction of $f_{\mu\nu}$. This behaviour is caused by the oscillation of the inflaton around its minimum, and implies that the dark sector metric undergoes multiple bounces during reheating. 

\subsection{The effects of reheating on tensor perturbations} \label{sub:newsec}

\begin{figure}
		\begin{center}
			\subfigure{\includegraphics[width=10cm]{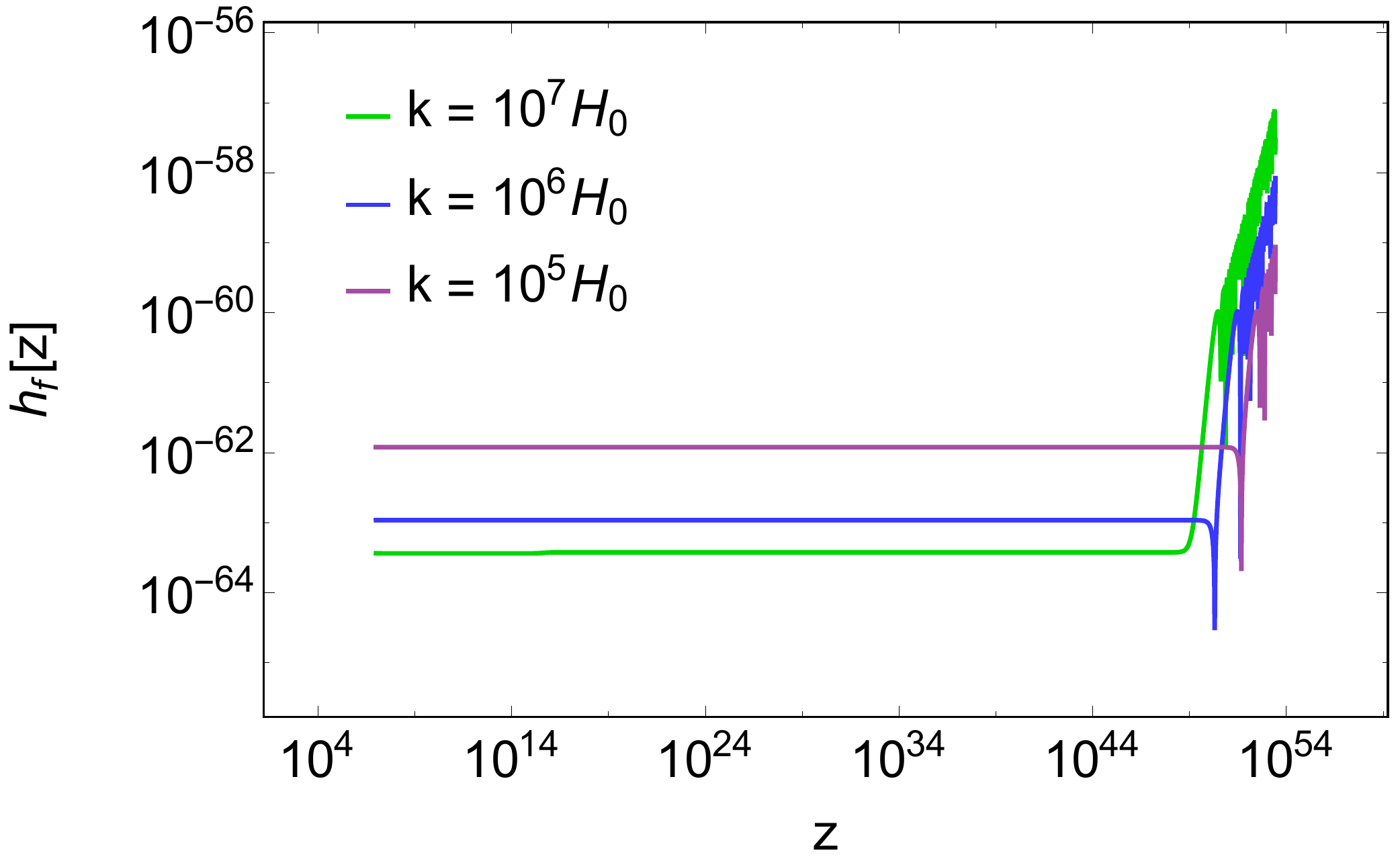}}
			\captionsetup{justification=raggedright,singlelinecheck=off,font=footnotesize}
			\caption{The evolution of $h_f$ from inflation (at redshift of $z \sim 10^{54}$), through reheating (at redshift $z \sim 10^{30}$), to the end of radiation domination (at redshift $z \sim 10^3$).}
			\label{fig:inflation_modes}
		\end{center}
\end{figure}

To determine the evolution of perturbations during the transition from inflation to a radiation dominated Universe, we have explicitly evolved the coupled evolution equations Eq.~\eqref{eq:tensorg} and~\eqref{eq:tensorf} in the background shown in Fig.~\ref{fig:inflation}. The evolution of a few modes from inflation, through reheating, to near the end of radiation domination, with initial conditions set by Eq.~\eqref{eq:mode_fns2}, is shown in Fig.~\ref{fig:inflation_modes}. On scales relevant for the CMB, we find that after horizon-crossing during inflation $h_f$ remains frozen through reheating and the duration of radiation domination. This validates the initial conditions assumed above, where only the constant mode of $h_f$ is excited at the beginning of radiation domination. On these large scales, we find good quantitative agreement with the prediction Eq.~\eqref{eq:velocity_condition}, and conclude inflationary initial conditions will yield no observable deviation from GR in the CMB.\\

On far smaller scales, of order the size of the horizon at reheating (in our model, this corresponds to $k \sim 10^{30} \mathcal{H}_0$), the growing mode is excited early in the radiation dominated era. For such modes, one should not apply the growth factor for $h_g$ found above (Eq.~\eqref{eq:growth}), but rather the growth factor found in Ref.~\cite{Cusin:2014aa}. Since this growth factor falls off like $k^4$, the growth in $h_g$ is highly suppressed, and the result will be no observable deviation from GR even though the growing mode dominates from the beginning of radiation domination.\\ 

\section{Conclusion} \label{sec:conclusion}

We have studied the properties of gravitational waves in massive bigravity, and their impact on cosmological observables compared to the standard predictions of General Relativity. The two background solutions we have studied display dramatically different phenomenology, illustrating the enormous size of the parameter space for observables. \\

In the ``expanding branch" in which both metrics expand in time, we found that for a wide range of initial conditions, the physical tensor perturbations $h_g$ matches closely with the pure GR solution. Due to a dramatic decay of $h_f$, the impact of the dark sector on $h_g$ is not important and causes no significant deviation from GR. \\

The ``bouncing branch", in which the dark metric $f_{\mu\nu}$ undergoes a bounce from contraction to expansion, has potentially dramatic differences from GR in the tensor sector. When the $f$ metric is undergoing contraction, the lapse $c$ is negative, which causes  $h_f$ to grow. This growth in $h_f$ can translate into growth in $h_g$ through the mixing term in the equations of motion, in some cases leading to physical gravitational waves with growing amplitudes at late times. This contrasts significantly with gravitational waves in GR which decay with time. The growth can potentially impact the CMB tensor power spectrum by dramatically amplifying large scale temperature anisotropies. The present day stochastic gravitational wave background, $\Omega_\text{GW}^0$ can be impacted as well, inheriting a very red spectrum that decays with the square of the wave number. The degree of growth depends on the scale of reheating, amplitude of the dark sector tensor modes, wave number, and graviton mass, and obeys the scaling relation Eq.~\ref{eq:growth} for initial amplitudes above a  critical value. On the largest scales, we find that the dark sector tensor modes can have a significant influence on the physical tensor for $h_f(\tau_i)/h_g(\tau_i) > 10^{-20}$ for a reheat temperature of $T_i=10^{10}$ GeV. In the absence of a theory of initial conditions, it is not clear that this holds. \\

To address the question of initial conditions, we computed the primordial power spectrum for dark and visible sector tensors in an inflationary cosmology. We found that the expanding branch is far beyond the perturbative regime, and therefore inaccessible to a semi-classical treatment. However, the primordial power spectra in the bouncing branch show that $h_f/h_g \sim H_0 / H^I_g \sim 10^{-57}$ for high scale inflation, and $h_f/h_g \sim 10^{-29}$ for low scale inflation. With this level of suppression, there will be no observable deviation from GR in the CMB or stochastic gravitational wave background. We presented an inflationary model that exhibits this explicitly. \\

Let us now discuss our results in the context of related work in the literature. While this work was in progress, Refs.~\cite{Cusin:2014aa,Amendola:2015tua} appeared with complementary investigations of tensor modes in bigravity. Ref. \cite{Cusin:2014aa} considered the bouncing branch chosen in this paper with identical parameters, but with an initial condition that was entirely composed of a growing mode of the dark sector tensor. However, as shown in Sec.~\ref{sec:initialconds} it appears that inflationary initial conditions do not excite the growing mode as significantly. In \cite{Amendola:2015tua} the authors considered varying the initial conditions in a phenomenological way, and specifically tested the effects on CMB spectra. Both of these works conclude that a tuning of the initial conditions can render the theory viable, which we show is actually possible in the inflationary paradigm. In this sense, our investigation is largely complementary to previous work; taken together, the range of possibilities is covered. Of course, another theory of
initial conditions may prevail, and a proper treatment of higher-order couplings between the scalar and tensor sectors may reveal a significant enhancement. However, in the context of linear perturbation theory in inflationary cosmology, it appears that the growing mode on large scales is not excited at the end of inflation.\\

It is clear that tensors can be a sensitive probe of massive bigravity. Looking to the future, the parameter space of nearly homogeneous solutions will soon be completely explored both at the level of the background and first order perturbations. In light of this, it is equally important to consider the theory for initial conditions in a broader sense, as illustrated by the strong dependence on initial conditions found in this and other papers. To this end, we plan to return to the question of inflationary model building in massive bigravity in future work. Scenarios with small but observable deviations from GR could serve as an important alternative hypothesis necessary for testing GR on cosmological scales and in future gravitational wave observatories. \\

\acknowledgements

We thank L. Boyle, K. Hinterbichler, and R. Ribeiro for helpful conversations. We thank  L. Amendola, M. Crisostomi, G. Cusin, R. Durrer, P. Guarato, F. K{\"o}nnig, M. Martinelli, M. Motta, V. Pettorino, and L. Pilo for comments on a draft of the paper. Research at Perimeter Institute is supported by the Government of Canada through Industry Canada and by the Province of Ontario through the Ministry of Research and Innovation. MCJ is supported by the National Science and Engineering Research Council through a Discovery grant. AT acknowledges support from the Vanier Canada Graduate Scholarship program.

\bibliography{bigrav_ref}  

\begin{thebibliography}{43}%
\makeatletter
\providecommand \@ifxundefined [1]{%
 \@ifx{#1\undefined}
}%
\providecommand \@ifnum [1]{%
 \ifnum #1\expandafter \@firstoftwo
 \else \expandafter \@secondoftwo
 \fi
}%
\providecommand \@ifx [1]{%
 \ifx #1\expandafter \@firstoftwo
 \else \expandafter \@secondoftwo
 \fi
}%
\providecommand \natexlab [1]{#1}%
\providecommand \enquote  [1]{``#1''}%
\providecommand \bibnamefont  [1]{#1}%
\providecommand \bibfnamefont [1]{#1}%
\providecommand \citenamefont [1]{#1}%
\providecommand \href@noop [0]{\@secondoftwo}%
\providecommand \href [0]{\begingroup \@sanitize@url \@href}%
\providecommand \@href[1]{\@@startlink{#1}\@@href}%
\providecommand \@@href[1]{\endgroup#1\@@endlink}%
\providecommand \@sanitize@url [0]{\catcode `\\12\catcode `\$12\catcode
  `\&12\catcode `\#12\catcode `\^12\catcode `\_12\catcode `\%12\relax}%
\providecommand \@@startlink[1]{}%
\providecommand \@@endlink[0]{}%
\providecommand \url  [0]{\begingroup\@sanitize@url \@url }%
\providecommand \@url [1]{\endgroup\@href {#1}{\urlprefix }}%
\providecommand \urlprefix  [0]{URL }%
\providecommand \Eprint [0]{\href }%
\providecommand \doibase [0]{http://dx.doi.org/}%
\providecommand \selectlanguage [0]{\@gobble}%
\providecommand \bibinfo  [0]{\@secondoftwo}%
\providecommand \bibfield  [0]{\@secondoftwo}%
\providecommand \translation [1]{[#1]}%
\providecommand \BibitemOpen [0]{}%
\providecommand \bibitemStop [0]{}%
\providecommand \bibitemNoStop [0]{.\EOS\space}%
\providecommand \EOS [0]{\spacefactor3000\relax}%
\providecommand \BibitemShut  [1]{\csname bibitem#1\endcsname}%
\let\auto@bib@innerbib\@empty
\bibitem [{\citenamefont {Fierz}\ and\ \citenamefont
  {Pauli}(1939)}]{Fierz:1939ix}%
  \BibitemOpen
  \bibfield  {author} {\bibinfo {author} {\bibfnamefont {M.}~\bibnamefont
  {Fierz}}\ and\ \bibinfo {author} {\bibfnamefont {W.}~\bibnamefont {Pauli}},\
  }\bibfield  {title} {\enquote {\bibinfo {title} {{On relativistic wave
  equations for particles of arbitrary spin in an electromagnetic field}},}\
  }\href {\doibase 10.1098/rspa.1939.0140} {\bibfield  {journal} {\bibinfo
  {journal} {Proc.Roy.Soc.Lond.}\ }\textbf {\bibinfo {volume} {A173}},\
  \bibinfo {pages} {211--232} (\bibinfo {year} {1939})}\BibitemShut {NoStop}%
\bibitem [{\citenamefont {de~Rham}\ \emph {et~al.}(2011)\citenamefont
  {de~Rham}, \citenamefont {Gabadadze},\ and\ \citenamefont
  {Tolley}}]{deRham:2010kj}%
  \BibitemOpen
  \bibfield  {author} {\bibinfo {author} {\bibfnamefont {Claudia}\ \bibnamefont
  {de~Rham}}, \bibinfo {author} {\bibfnamefont {Gregory}\ \bibnamefont
  {Gabadadze}}, \ and\ \bibinfo {author} {\bibfnamefont {Andrew~J.}\
  \bibnamefont {Tolley}},\ }\bibfield  {title} {\enquote {\bibinfo {title}
  {{Resummation of Massive Gravity}},}\ }\href {\doibase
  10.1103/PhysRevLett.106.231101} {\bibfield  {journal} {\bibinfo  {journal}
  {Phys.Rev.Lett.}\ }\textbf {\bibinfo {volume} {106}},\ \bibinfo {pages}
  {231101} (\bibinfo {year} {2011})},\ \Eprint {http://arxiv.org/abs/1011.1232}
  {arXiv:1011.1232 [hep-th]} \BibitemShut {NoStop}%
\bibitem [{\citenamefont {de~Rham}(2014)}]{deRham:2014zqa}%
  \BibitemOpen
  \bibfield  {author} {\bibinfo {author} {\bibfnamefont {Claudia}\ \bibnamefont
  {de~Rham}},\ }\bibfield  {title} {\enquote {\bibinfo {title} {{Massive
  Gravity}},}\ }\href@noop {} {\  (\bibinfo {year} {2014})},\ \Eprint
  {http://arxiv.org/abs/1401.4173} {arXiv:1401.4173 [hep-th]} \BibitemShut
  {NoStop}%
\bibitem [{\citenamefont {{Hassan}}\ and\ \citenamefont
  {{Rosen}}(2012)}]{Hassan:2012aa}%
  \BibitemOpen
  \bibfield  {author} {\bibinfo {author} {\bibfnamefont {S.~F.}\ \bibnamefont
  {{Hassan}}}\ and\ \bibinfo {author} {\bibfnamefont {R.~A.}\ \bibnamefont
  {{Rosen}}},\ }\bibfield  {title} {\enquote {\bibinfo {title} {{Bimetric
  gravity from ghost-free massive gravity}},}\ }\href {\doibase
  10.1007/JHEP02(2012)126} {\bibfield  {journal} {\bibinfo  {journal} {Journal
  of High Energy Physics}\ }\textbf {\bibinfo {volume} {2}},\ \bibinfo {eid}
  {126} (\bibinfo {year} {2012})},\ \Eprint {http://arxiv.org/abs/1109.3515}
  {arXiv:1109.3515 [hep-th]} \BibitemShut {NoStop}%
\bibitem [{\citenamefont {{de Rham}}\ \emph {et~al.}(2014)\citenamefont {{de
  Rham}}, \citenamefont {{Heisenberg}},\ and\ \citenamefont
  {{Ribeiro}}}]{de-Rham:2014aa}%
  \BibitemOpen
  \bibfield  {author} {\bibinfo {author} {\bibfnamefont {C.}~\bibnamefont {{de
  Rham}}}, \bibinfo {author} {\bibfnamefont {L.}~\bibnamefont {{Heisenberg}}},
  \ and\ \bibinfo {author} {\bibfnamefont {R.~H.}\ \bibnamefont {{Ribeiro}}},\
  }\bibfield  {title} {\enquote {\bibinfo {title} {{On couplings to matter in
  massive (bi-)gravity}},}\ }\href@noop {} {\bibfield  {journal} {\bibinfo
  {journal} {ArXiv e-prints}\ } (\bibinfo {year} {2014})},\ \Eprint
  {http://arxiv.org/abs/1408.1678} {arXiv:1408.1678 [hep-th]} \BibitemShut
  {NoStop}%
\bibitem [{\citenamefont {{K{\"o}nnig}}\ \emph
  {et~al.}(2014{\natexlab{a}})\citenamefont {{K{\"o}nnig}}, \citenamefont
  {{Patil}},\ and\ \citenamefont {{Amendola}}}]{Koennig:2014ab}%
  \BibitemOpen
  \bibfield  {author} {\bibinfo {author} {\bibfnamefont {F.}~\bibnamefont
  {{K{\"o}nnig}}}, \bibinfo {author} {\bibfnamefont {A.}~\bibnamefont
  {{Patil}}}, \ and\ \bibinfo {author} {\bibfnamefont {L.}~\bibnamefont
  {{Amendola}}},\ }\bibfield  {title} {\enquote {\bibinfo {title} {{Viable
  cosmological solutions in massive bimetric gravity}},}\ }\href {\doibase
  10.1088/1475-7516/2014/03/029} {\bibfield  {journal} {\bibinfo  {journal}
  {jcap}\ }\textbf {\bibinfo {volume} {3}},\ \bibinfo {eid} {029} (\bibinfo
  {year} {2014}{\natexlab{a}})},\ \Eprint {http://arxiv.org/abs/1312.3208}
  {arXiv:1312.3208 [astro-ph.CO]} \BibitemShut {NoStop}%
\bibitem [{\citenamefont {{Akrami}}\ \emph {et~al.}(2013)\citenamefont
  {{Akrami}}, \citenamefont {{Koivisto}},\ and\ \citenamefont
  {{Sandstad}}}]{Akrami:2013ab}%
  \BibitemOpen
  \bibfield  {author} {\bibinfo {author} {\bibfnamefont {Y.}~\bibnamefont
  {{Akrami}}}, \bibinfo {author} {\bibfnamefont {T.~S.}\ \bibnamefont
  {{Koivisto}}}, \ and\ \bibinfo {author} {\bibfnamefont {M.}~\bibnamefont
  {{Sandstad}}},\ }\bibfield  {title} {\enquote {\bibinfo {title}
  {{Cosmological constraints on ghost-free bigravity: background dynamics and
  late-time acceleration}},}\ }\href@noop {} {\bibfield  {journal} {\bibinfo
  {journal} {ArXiv e-prints}\ } (\bibinfo {year} {2013})},\ \Eprint
  {http://arxiv.org/abs/1302.5268} {arXiv:1302.5268 [astro-ph.CO]} \BibitemShut
  {NoStop}%
\bibitem [{\citenamefont {{von Strauss}}\ \emph {et~al.}(2012)\citenamefont
  {{von Strauss}}, \citenamefont {{Schmidt-May}}, \citenamefont {{Enander}},
  \citenamefont {{M{\"o}rtsell}},\ and\ \citenamefont
  {{Hassan}}}]{von-Strauss:2012aa}%
  \BibitemOpen
  \bibfield  {author} {\bibinfo {author} {\bibfnamefont {M.}~\bibnamefont {{von
  Strauss}}}, \bibinfo {author} {\bibfnamefont {A.}~\bibnamefont
  {{Schmidt-May}}}, \bibinfo {author} {\bibfnamefont {J.}~\bibnamefont
  {{Enander}}}, \bibinfo {author} {\bibfnamefont {E.}~\bibnamefont
  {{M{\"o}rtsell}}}, \ and\ \bibinfo {author} {\bibfnamefont {S.~F.}\
  \bibnamefont {{Hassan}}},\ }\bibfield  {title} {\enquote {\bibinfo {title}
  {{Cosmological solutions in bimetric gravity and their observational
  tests}},}\ }\href {\doibase 10.1088/1475-7516/2012/03/042} {\bibfield
  {journal} {\bibinfo  {journal} {jcap}\ }\textbf {\bibinfo {volume} {3}},\
  \bibinfo {eid} {042} (\bibinfo {year} {2012})},\ \Eprint
  {http://arxiv.org/abs/1111.1655} {arXiv:1111.1655 [gr-qc]} \BibitemShut
  {NoStop}%
\bibitem [{\citenamefont {{Tamanini}}\ \emph {et~al.}(2014)\citenamefont
  {{Tamanini}}, \citenamefont {{Saridakis}},\ and\ \citenamefont
  {{Koivisto}}}]{Tamanini:2014aa}%
  \BibitemOpen
  \bibfield  {author} {\bibinfo {author} {\bibfnamefont {N.}~\bibnamefont
  {{Tamanini}}}, \bibinfo {author} {\bibfnamefont {E.~N.}\ \bibnamefont
  {{Saridakis}}}, \ and\ \bibinfo {author} {\bibfnamefont {T.~S.}\ \bibnamefont
  {{Koivisto}}},\ }\bibfield  {title} {\enquote {\bibinfo {title} {{The
  cosmology of interacting spin-2 fields}},}\ }\href {\doibase
  10.1088/1475-7516/2014/02/015} {\bibfield  {journal} {\bibinfo  {journal}
  {JCAP}\ }\textbf {\bibinfo {volume} {2}},\ \bibinfo {eid} {015} (\bibinfo
  {year} {2014})},\ \Eprint {http://arxiv.org/abs/1307.5984} {arXiv:1307.5984
  [hep-th]} \BibitemShut {NoStop}%
\bibitem [{\citenamefont {{Comelli}}\ \emph
  {et~al.}(2012{\natexlab{a}})\citenamefont {{Comelli}}, \citenamefont
  {{Crisostomi}}, \citenamefont {{Nesti}},\ and\ \citenamefont
  {{Pilo}}}]{Comelli:2012ab}%
  \BibitemOpen
  \bibfield  {author} {\bibinfo {author} {\bibfnamefont {D.}~\bibnamefont
  {{Comelli}}}, \bibinfo {author} {\bibfnamefont {M.}~\bibnamefont
  {{Crisostomi}}}, \bibinfo {author} {\bibfnamefont {F.}~\bibnamefont
  {{Nesti}}}, \ and\ \bibinfo {author} {\bibfnamefont {L.}~\bibnamefont
  {{Pilo}}},\ }\bibfield  {title} {\enquote {\bibinfo {title} {{FRW cosmology
  in ghost free massive gravity from bigravity}},}\ }\href {\doibase
  10.1007/JHEP03(2012)067} {\bibfield  {journal} {\bibinfo  {journal} {Journal
  of High Energy Physics}\ }\textbf {\bibinfo {volume} {3}},\ \bibinfo {eid}
  {67} (\bibinfo {year} {2012}{\natexlab{a}})},\ \Eprint
  {http://arxiv.org/abs/1111.1983} {arXiv:1111.1983 [hep-th]} \BibitemShut
  {NoStop}%
\bibitem [{\citenamefont {{Volkov}}(2012)}]{Volkov:2012aa}%
  \BibitemOpen
  \bibfield  {author} {\bibinfo {author} {\bibfnamefont {M.~S.}\ \bibnamefont
  {{Volkov}}},\ }\bibfield  {title} {\enquote {\bibinfo {title} {{Cosmological
  solutions with massive gravitons in the bigravity theory}},}\ }\href
  {\doibase 10.1007/JHEP01(2012)035} {\bibfield  {journal} {\bibinfo  {journal}
  {Journal of High Energy Physics}\ }\textbf {\bibinfo {volume} {1}},\ \bibinfo
  {eid} {35} (\bibinfo {year} {2012})},\ \Eprint
  {http://arxiv.org/abs/1110.6153} {arXiv:1110.6153 [hep-th]} \BibitemShut
  {NoStop}%
\bibitem [{\citenamefont {{Solomon}}\ \emph {et~al.}(2014)\citenamefont
  {{Solomon}}, \citenamefont {{Akrami}},\ and\ \citenamefont
  {{Koivisto}}}]{Solomon:2014aa}%
  \BibitemOpen
  \bibfield  {author} {\bibinfo {author} {\bibfnamefont {A.~R.}\ \bibnamefont
  {{Solomon}}}, \bibinfo {author} {\bibfnamefont {Y.}~\bibnamefont {{Akrami}}},
  \ and\ \bibinfo {author} {\bibfnamefont {T.~S.}\ \bibnamefont {{Koivisto}}},\
  }\bibfield  {title} {\enquote {\bibinfo {title} {{Linear growth of structure
  in massive bigravity}},}\ }\href@noop {} {\bibfield  {journal} {\bibinfo
  {journal} {ArXiv e-prints}\ } (\bibinfo {year} {2014})},\ \Eprint
  {http://arxiv.org/abs/1404.4061} {arXiv:1404.4061} \BibitemShut {NoStop}%
\bibitem [{\citenamefont {{Comelli}}\ \emph
  {et~al.}(2012{\natexlab{b}})\citenamefont {{Comelli}}, \citenamefont
  {{Crisostomi}},\ and\ \citenamefont {{Pilo}}}]{Comelli:2012aa}%
  \BibitemOpen
  \bibfield  {author} {\bibinfo {author} {\bibfnamefont {D.}~\bibnamefont
  {{Comelli}}}, \bibinfo {author} {\bibfnamefont {M.}~\bibnamefont
  {{Crisostomi}}}, \ and\ \bibinfo {author} {\bibfnamefont {L.}~\bibnamefont
  {{Pilo}}},\ }\bibfield  {title} {\enquote {\bibinfo {title} {{Perturbations
  in massive gravity cosmology}},}\ }\href {\doibase 10.1007/JHEP06(2012)085}
  {\bibfield  {journal} {\bibinfo  {journal} {Journal of High Energy Physics}\
  }\textbf {\bibinfo {volume} {6}},\ \bibinfo {eid} {85} (\bibinfo {year}
  {2012}{\natexlab{b}})},\ \Eprint {http://arxiv.org/abs/1202.1986}
  {arXiv:1202.1986 [hep-th]} \BibitemShut {NoStop}%
\bibitem [{\citenamefont {{K{\"o}nnig}}\ and\ \citenamefont
  {{Amendola}}(2014{\natexlab{a}})}]{Koennig:2014aa}%
  \BibitemOpen
  \bibfield  {author} {\bibinfo {author} {\bibfnamefont {F.}~\bibnamefont
  {{K{\"o}nnig}}}\ and\ \bibinfo {author} {\bibfnamefont {L.}~\bibnamefont
  {{Amendola}}},\ }\bibfield  {title} {\enquote {\bibinfo {title} {{Instability
  in a minimal bimetric gravity model}},}\ }\href {\doibase
  10.1103/PhysRevD.90.044030} {\bibfield  {journal} {\bibinfo  {journal}
  {\prd}\ }\textbf {\bibinfo {volume} {90}},\ \bibinfo {eid} {044030} (\bibinfo
  {year} {2014}{\natexlab{a}})},\ \Eprint {http://arxiv.org/abs/1402.1988v2}
  {arXiv:1402.1988v2} \BibitemShut {NoStop}%
\bibitem [{\citenamefont {{Berg}}\ \emph {et~al.}(2012)\citenamefont {{Berg}},
  \citenamefont {{Buchberger}}, \citenamefont {{Enander}}, \citenamefont
  {{M{\"o}rtsell}},\ and\ \citenamefont {{Sj{\"o}rs}}}]{Berg:2012aa}%
  \BibitemOpen
  \bibfield  {author} {\bibinfo {author} {\bibfnamefont {M.}~\bibnamefont
  {{Berg}}}, \bibinfo {author} {\bibfnamefont {I.}~\bibnamefont
  {{Buchberger}}}, \bibinfo {author} {\bibfnamefont {J.}~\bibnamefont
  {{Enander}}}, \bibinfo {author} {\bibfnamefont {E.}~\bibnamefont
  {{M{\"o}rtsell}}}, \ and\ \bibinfo {author} {\bibfnamefont {S.}~\bibnamefont
  {{Sj{\"o}rs}}},\ }\bibfield  {title} {\enquote {\bibinfo {title} {{Growth
  histories in bimetric massive gravity}},}\ }\href {\doibase
  10.1088/1475-7516/2012/12/021} {\bibfield  {journal} {\bibinfo  {journal}
  {jcap}\ }\textbf {\bibinfo {volume} {12}},\ \bibinfo {eid} {021} (\bibinfo
  {year} {2012})},\ \Eprint {http://arxiv.org/abs/1206.3496} {arXiv:1206.3496
  [gr-qc]} \BibitemShut {NoStop}%
\bibitem [{\citenamefont {{Lagos}}\ and\ \citenamefont
  {{Ferreira}}(2014)}]{Lagos:2014aa}%
  \BibitemOpen
  \bibfield  {author} {\bibinfo {author} {\bibfnamefont {M.}~\bibnamefont
  {{Lagos}}}\ and\ \bibinfo {author} {\bibfnamefont {P.~G.}\ \bibnamefont
  {{Ferreira}}},\ }\bibfield  {title} {\enquote {\bibinfo {title}
  {{Cosmological perturbations in massive bigravity}},}\ }\href@noop {}
  {\bibfield  {journal} {\bibinfo  {journal} {ArXiv e-prints}\ } (\bibinfo
  {year} {2014})},\ \Eprint {http://arxiv.org/abs/1410.0207} {arXiv:1410.0207
  [gr-qc]} \BibitemShut {NoStop}%
\bibitem [{\citenamefont {{Cusin}}\ \emph {et~al.}(2014)\citenamefont
  {{Cusin}}, \citenamefont {{Durrer}}, \citenamefont {{Guarato}},\ and\
  \citenamefont {{Motta}}}]{Cusin:2014aa}%
  \BibitemOpen
  \bibfield  {author} {\bibinfo {author} {\bibfnamefont {G.}~\bibnamefont
  {{Cusin}}}, \bibinfo {author} {\bibfnamefont {R.}~\bibnamefont {{Durrer}}},
  \bibinfo {author} {\bibfnamefont {P.}~\bibnamefont {{Guarato}}}, \ and\
  \bibinfo {author} {\bibfnamefont {M.}~\bibnamefont {{Motta}}},\ }\bibfield
  {title} {\enquote {\bibinfo {title} {{Gravitational waves in bigravity
  cosmology}},}\ }\href@noop {} {\bibfield  {journal} {\bibinfo  {journal}
  {ArXiv e-prints}\ } (\bibinfo {year} {2014})},\ \Eprint
  {http://arxiv.org/abs/1412.5979} {arXiv:1412.5979} \BibitemShut {NoStop}%
\bibitem [{\citenamefont {De~Felice}\ \emph {et~al.}(2014)\citenamefont
  {De~Felice}, \citenamefont {Nakamura},\ and\ \citenamefont
  {Tanaka}}]{DeFelice:2013nba}%
  \BibitemOpen
  \bibfield  {author} {\bibinfo {author} {\bibfnamefont {Antonio}\ \bibnamefont
  {De~Felice}}, \bibinfo {author} {\bibfnamefont {Takashi}\ \bibnamefont
  {Nakamura}}, \ and\ \bibinfo {author} {\bibfnamefont {Takahiro}\ \bibnamefont
  {Tanaka}},\ }\bibfield  {title} {\enquote {\bibinfo {title} {{Possible
  existence of viable models of bi-gravity with detectable graviton
  oscillations by gravitational wave detectors}},}\ }\href {\doibase
  10.1093/ptep/ptu024} {\bibfield  {journal} {\bibinfo  {journal} {PTEP}\
  }\textbf {\bibinfo {volume} {2014}},\ \bibinfo {pages} {043E01} (\bibinfo
  {year} {2014})},\ \Eprint {http://arxiv.org/abs/1304.3920} {arXiv:1304.3920
  [gr-qc]} \BibitemShut {NoStop}%
\bibitem [{\citenamefont {Brito}\ \emph {et~al.}(2013)\citenamefont {Brito},
  \citenamefont {Cardoso},\ and\ \citenamefont {Pani}}]{Brito:2013wya}%
  \BibitemOpen
  \bibfield  {author} {\bibinfo {author} {\bibfnamefont {Richard}\ \bibnamefont
  {Brito}}, \bibinfo {author} {\bibfnamefont {Vitor}\ \bibnamefont {Cardoso}},
  \ and\ \bibinfo {author} {\bibfnamefont {Paolo}\ \bibnamefont {Pani}},\
  }\bibfield  {title} {\enquote {\bibinfo {title} {{Massive spin-2 fields on
  black hole spacetimes: Instability of the Schwarzschild and Kerr solutions
  and bounds on the graviton mass}},}\ }\href {\doibase
  10.1103/PhysRevD.88.023514} {\bibfield  {journal} {\bibinfo  {journal}
  {Phys.Rev.}\ }\textbf {\bibinfo {volume} {D88}},\ \bibinfo {pages} {023514}
  (\bibinfo {year} {2013})},\ \Eprint {http://arxiv.org/abs/1304.6725}
  {arXiv:1304.6725 [gr-qc]} \BibitemShut {NoStop}%
\bibitem [{\citenamefont {Saltas}\ \emph {et~al.}(2014)\citenamefont {Saltas},
  \citenamefont {Sawicki}, \citenamefont {Amendola},\ and\ \citenamefont
  {Kunz}}]{Saltas:2014dha}%
  \BibitemOpen
  \bibfield  {author} {\bibinfo {author} {\bibfnamefont {Ippocratis~D.}\
  \bibnamefont {Saltas}}, \bibinfo {author} {\bibfnamefont {Ignacy}\
  \bibnamefont {Sawicki}}, \bibinfo {author} {\bibfnamefont {Luca}\
  \bibnamefont {Amendola}}, \ and\ \bibinfo {author} {\bibfnamefont {Martin}\
  \bibnamefont {Kunz}},\ }\bibfield  {title} {\enquote {\bibinfo {title}
  {{Anisotropic Stress as a Signature of Nonstandard Propagation of
  Gravitational Waves}},}\ }\href {\doibase 10.1103/PhysRevLett.113.191101}
  {\bibfield  {journal} {\bibinfo  {journal} {Phys.Rev.Lett.}\ }\textbf
  {\bibinfo {volume} {113}},\ \bibinfo {pages} {191101} (\bibinfo {year}
  {2014})},\ \Eprint {http://arxiv.org/abs/1406.7139} {arXiv:1406.7139
  [astro-ph.CO]} \BibitemShut {NoStop}%
\bibitem [{\citenamefont {{Berezhiani}}\ \emph {et~al.}(2007)\citenamefont
  {{Berezhiani}}, \citenamefont {{Comelli}}, \citenamefont {{Nesti}},\ and\
  \citenamefont {{Pilo}}}]{Berezhiani:2007aa}%
  \BibitemOpen
  \bibfield  {author} {\bibinfo {author} {\bibfnamefont {Z.}~\bibnamefont
  {{Berezhiani}}}, \bibinfo {author} {\bibfnamefont {D.}~\bibnamefont
  {{Comelli}}}, \bibinfo {author} {\bibfnamefont {F.}~\bibnamefont {{Nesti}}},
  \ and\ \bibinfo {author} {\bibfnamefont {L.}~\bibnamefont {{Pilo}}},\
  }\bibfield  {title} {\enquote {\bibinfo {title} {{Spontaneous Lorentz
  Breaking and Massive Gravity}},}\ }\href {\doibase
  10.1103/PhysRevLett.99.131101} {\bibfield  {journal} {\bibinfo  {journal}
  {Physical Review Letters}\ }\textbf {\bibinfo {volume} {99}},\ \bibinfo {eid}
  {131101} (\bibinfo {year} {2007})},\ \Eprint
  {http://arxiv.org/abs/hep-th/0703264} {hep-th/0703264} \BibitemShut {NoStop}%
\bibitem [{\citenamefont {Raveri}\ \emph {et~al.}(2014)\citenamefont {Raveri},
  \citenamefont {Baccigalupi}, \citenamefont {Silvestri},\ and\ \citenamefont
  {Zhou}}]{Raveri:2014eea}%
  \BibitemOpen
  \bibfield  {author} {\bibinfo {author} {\bibfnamefont {Marco}\ \bibnamefont
  {Raveri}}, \bibinfo {author} {\bibfnamefont {Carlo}\ \bibnamefont
  {Baccigalupi}}, \bibinfo {author} {\bibfnamefont {Alessandra}\ \bibnamefont
  {Silvestri}}, \ and\ \bibinfo {author} {\bibfnamefont {Shuang-Yong}\
  \bibnamefont {Zhou}},\ }\bibfield  {title} {\enquote {\bibinfo {title}
  {{Measuring the speed of cosmological gravitational waves}},}\ }\href@noop {}
  {\  (\bibinfo {year} {2014})},\ \Eprint {http://arxiv.org/abs/1405.7974}
  {arXiv:1405.7974 [astro-ph.CO]} \BibitemShut {NoStop}%
\bibitem [{\citenamefont {Xu}(2014)}]{Xu:2014uba}%
  \BibitemOpen
  \bibfield  {author} {\bibinfo {author} {\bibfnamefont {Lixin}\ \bibnamefont
  {Xu}},\ }\bibfield  {title} {\enquote {\bibinfo {title} {{Gravitational
  Waves: A Test for Modified Gravity}},}\ }\href@noop {} {\  (\bibinfo {year}
  {2014})},\ \Eprint {http://arxiv.org/abs/1410.6977} {arXiv:1410.6977
  [astro-ph.CO]} \BibitemShut {NoStop}%
\bibitem [{\citenamefont {Amendola}\ \emph {et~al.}(2014)\citenamefont
  {Amendola}, \citenamefont {Ballesteros},\ and\ \citenamefont
  {Pettorino}}]{Amendola:2014wma}%
  \BibitemOpen
  \bibfield  {author} {\bibinfo {author} {\bibfnamefont {Luca}\ \bibnamefont
  {Amendola}}, \bibinfo {author} {\bibfnamefont {Guillermo}\ \bibnamefont
  {Ballesteros}}, \ and\ \bibinfo {author} {\bibfnamefont {Valeria}\
  \bibnamefont {Pettorino}},\ }\bibfield  {title} {\enquote {\bibinfo {title}
  {{Effects of modified gravity on B-mode polarization}},}\ }\href {\doibase
  10.1103/PhysRevD.90.043009} {\bibfield  {journal} {\bibinfo  {journal}
  {Phys.Rev.}\ }\textbf {\bibinfo {volume} {D90}},\ \bibinfo {pages} {043009}
  (\bibinfo {year} {2014})},\ \Eprint {http://arxiv.org/abs/1405.7004}
  {arXiv:1405.7004 [astro-ph.CO]} \BibitemShut {NoStop}%
\bibitem [{\citenamefont {Pettorino}\ and\ \citenamefont
  {Amendola}(2015)}]{Pettorino:2014bka}%
  \BibitemOpen
  \bibfield  {author} {\bibinfo {author} {\bibfnamefont {Valeria}\ \bibnamefont
  {Pettorino}}\ and\ \bibinfo {author} {\bibfnamefont {Luca}\ \bibnamefont
  {Amendola}},\ }\bibfield  {title} {\enquote {\bibinfo {title} {{Friction in
  Gravitational Waves: a test for early-time modified gravity}},}\ }\href
  {\doibase 10.1016/j.physletb.2015.02.007} {\bibfield  {journal} {\bibinfo
  {journal} {Phys.Lett.}\ }\textbf {\bibinfo {volume} {B742}},\ \bibinfo
  {pages} {353--357} (\bibinfo {year} {2015})},\ \Eprint
  {http://arxiv.org/abs/1408.2224} {arXiv:1408.2224 [astro-ph.CO]} \BibitemShut
  {NoStop}%
\bibitem [{\citenamefont {Cai}\ \emph {et~al.}(2015)\citenamefont {Cai},
  \citenamefont {Wang},\ and\ \citenamefont {Piao}}]{Cai:2015dta}%
  \BibitemOpen
  \bibfield  {author} {\bibinfo {author} {\bibfnamefont {Yong}\ \bibnamefont
  {Cai}}, \bibinfo {author} {\bibfnamefont {Yu-Tong}\ \bibnamefont {Wang}}, \
  and\ \bibinfo {author} {\bibfnamefont {Yun-Song}\ \bibnamefont {Piao}},\
  }\bibfield  {title} {\enquote {\bibinfo {title} {{Oscillating modulation to
  B-mode polarization from varying propagating speed of primordial
  gravitational waves}},}\ }\href@noop {} {\  (\bibinfo {year} {2015})},\
  \Eprint {http://arxiv.org/abs/1501.06345} {arXiv:1501.06345 [astro-ph.CO]}
  \BibitemShut {NoStop}%
\bibitem [{\citenamefont {Amendola}\ \emph {et~al.}(2015)\citenamefont
  {Amendola}, \citenamefont {K{\"o}nnig}, \citenamefont {Martinelli},
  \citenamefont {Pettorino},\ and\ \citenamefont
  {Zumalacarregui}}]{Amendola:2015tua}%
  \BibitemOpen
  \bibfield  {author} {\bibinfo {author} {\bibfnamefont {Luca}\ \bibnamefont
  {Amendola}}, \bibinfo {author} {\bibfnamefont {Frank}\ \bibnamefont
  {K{\"o}nnig}}, \bibinfo {author} {\bibfnamefont {Matteo}\ \bibnamefont
  {Martinelli}}, \bibinfo {author} {\bibfnamefont {Valeria}\ \bibnamefont
  {Pettorino}}, \ and\ \bibinfo {author} {\bibfnamefont {Miguel}\ \bibnamefont
  {Zumalacarregui}},\ }\bibfield  {title} {\enquote {\bibinfo {title} {{Surfing
  gravitational waves: can bigravity survive growing tensor modes?}}}\
  }\href@noop {} {\  (\bibinfo {year} {2015})},\ \Eprint
  {http://arxiv.org/abs/1503.02490} {arXiv:1503.02490 [astro-ph.CO]}
  \BibitemShut {NoStop}%
\bibitem [{\citenamefont {{Comelli}}\ \emph
  {et~al.}(2012{\natexlab{c}})\citenamefont {{Comelli}}, \citenamefont
  {{Crisostomi}}, \citenamefont {{Nesti}},\ and\ \citenamefont
  {{Pilo}}}]{Comelli:2012ac}%
  \BibitemOpen
  \bibfield  {author} {\bibinfo {author} {\bibfnamefont {D.}~\bibnamefont
  {{Comelli}}}, \bibinfo {author} {\bibfnamefont {M.}~\bibnamefont
  {{Crisostomi}}}, \bibinfo {author} {\bibfnamefont {F.}~\bibnamefont
  {{Nesti}}}, \ and\ \bibinfo {author} {\bibfnamefont {L.}~\bibnamefont
  {{Pilo}}},\ }\bibfield  {title} {\enquote {\bibinfo {title} {{Spherically
  symmetric solutions in ghost-free massive gravity}},}\ }\href {\doibase
  10.1103/PhysRevD.85.024044} {\bibfield  {journal} {\bibinfo  {journal}
  {\prd}\ }\textbf {\bibinfo {volume} {85}},\ \bibinfo {eid} {024044} (\bibinfo
  {year} {2012}{\natexlab{c}})},\ \Eprint {http://arxiv.org/abs/1110.4967}
  {arXiv:1110.4967 [hep-th]} \BibitemShut {NoStop}%
\bibitem [{\citenamefont {{K{\"o}nnig}}\ and\ \citenamefont
  {{Amendola}}(2014{\natexlab{b}})}]{Koennig:2014aav1}%
  \BibitemOpen
  \bibfield  {author} {\bibinfo {author} {\bibfnamefont {F.}~\bibnamefont
  {{K{\"o}nnig}}}\ and\ \bibinfo {author} {\bibfnamefont {L.}~\bibnamefont
  {{Amendola}}},\ }\bibfield  {title} {\enquote {\bibinfo {title} {{A minimal
  bimetric gravity model that fits cosmological observations}},}\ }\href@noop
  {} {\bibfield  {journal} {\bibinfo  {journal} {ArXiv e-prints}\ ,\ \bibinfo
  {eid} {044030}} (\bibinfo {year} {2014}{\natexlab{b}})},\ \Eprint
  {http://arxiv.org/abs/1402.1988v1} {arXiv:1402.1988v1} \BibitemShut {NoStop}%
\bibitem [{\citenamefont {{K{\"o}nnig}}\ \emph
  {et~al.}(2014{\natexlab{b}})\citenamefont {{K{\"o}nnig}}, \citenamefont
  {{Akrami}}, \citenamefont {{Amendola}}, \citenamefont {{Motta}},\ and\
  \citenamefont {{Solomon}}}]{Konnig:2014aa}%
  \BibitemOpen
  \bibfield  {author} {\bibinfo {author} {\bibfnamefont {F.}~\bibnamefont
  {{K{\"o}nnig}}}, \bibinfo {author} {\bibfnamefont {Y.}~\bibnamefont
  {{Akrami}}}, \bibinfo {author} {\bibfnamefont {L.}~\bibnamefont
  {{Amendola}}}, \bibinfo {author} {\bibfnamefont {M.}~\bibnamefont {{Motta}}},
  \ and\ \bibinfo {author} {\bibfnamefont {A.~R.}\ \bibnamefont {{Solomon}}},\
  }\bibfield  {title} {\enquote {\bibinfo {title} {{Stable and unstable
  cosmological models in bimetric massive gravity}},}\ }\href@noop {}
  {\bibfield  {journal} {\bibinfo  {journal} {ArXiv e-prints}\ } (\bibinfo
  {year} {2014}{\natexlab{b}})},\ \Eprint {http://arxiv.org/abs/1407.4331}
  {arXiv:1407.4331} \BibitemShut {NoStop}%
\bibitem [{\citenamefont {{Gratia}}\ \emph {et~al.}(2013)\citenamefont
  {{Gratia}}, \citenamefont {{Hu}},\ and\ \citenamefont
  {{Wyman}}}]{Gratia:2013aa}%
  \BibitemOpen
  \bibfield  {author} {\bibinfo {author} {\bibfnamefont {P.}~\bibnamefont
  {{Gratia}}}, \bibinfo {author} {\bibfnamefont {W.}~\bibnamefont {{Hu}}}, \
  and\ \bibinfo {author} {\bibfnamefont {M.}~\bibnamefont {{Wyman}}},\
  }\bibfield  {title} {\enquote {\bibinfo {title} {{Self-accelerating massive
  gravity: how zweibeins walk through determinant singularities}},}\ }\href
  {\doibase 10.1088/0264-9381/30/18/184007} {\bibfield  {journal} {\bibinfo
  {journal} {Classical and Quantum Gravity}\ }\textbf {\bibinfo {volume}
  {30}},\ \bibinfo {eid} {184007} (\bibinfo {year} {2013})},\ \Eprint
  {http://arxiv.org/abs/1305.2916} {arXiv:1305.2916 [hep-th]} \BibitemShut
  {NoStop}%
\bibitem [{\citenamefont {{Gratia}}\ \emph {et~al.}(2014)\citenamefont
  {{Gratia}}, \citenamefont {{Hu}},\ and\ \citenamefont
  {{Wyman}}}]{Gratia:2014aa}%
  \BibitemOpen
  \bibfield  {author} {\bibinfo {author} {\bibfnamefont {P.}~\bibnamefont
  {{Gratia}}}, \bibinfo {author} {\bibfnamefont {W.}~\bibnamefont {{Hu}}}, \
  and\ \bibinfo {author} {\bibfnamefont {M.}~\bibnamefont {{Wyman}}},\
  }\bibfield  {title} {\enquote {\bibinfo {title} {{Self-accelerating massive
  gravity: Bimetric determinant singularities}},}\ }\href {\doibase
  10.1103/PhysRevD.89.027502} {\bibfield  {journal} {\bibinfo  {journal}
  {\prd}\ }\textbf {\bibinfo {volume} {89}},\ \bibinfo {eid} {027502} (\bibinfo
  {year} {2014})},\ \Eprint {http://arxiv.org/abs/1309.5947} {arXiv:1309.5947
  [hep-th]} \BibitemShut {NoStop}%
\bibitem [{\citenamefont {{Enander}}\ \emph {et~al.}(2015)\citenamefont
  {{Enander}}, \citenamefont {{Akrami}}, \citenamefont {{Mortsell}},
  \citenamefont {{Renneby}},\ and\ \citenamefont {{Solomon}}}]{Enander:2015aa}%
  \BibitemOpen
  \bibfield  {author} {\bibinfo {author} {\bibfnamefont {J.}~\bibnamefont
  {{Enander}}}, \bibinfo {author} {\bibfnamefont {Y.}~\bibnamefont {{Akrami}}},
  \bibinfo {author} {\bibfnamefont {E.}~\bibnamefont {{Mortsell}}}, \bibinfo
  {author} {\bibfnamefont {M.}~\bibnamefont {{Renneby}}}, \ and\ \bibinfo
  {author} {\bibfnamefont {A.~R.}\ \bibnamefont {{Solomon}}},\ }\bibfield
  {title} {\enquote {\bibinfo {title} {{Integrated Sachs-Wolfe effect in
  massive bigravity}},}\ }\href@noop {} {\bibfield  {journal} {\bibinfo
  {journal} {ArXiv e-prints}\ } (\bibinfo {year} {2015})},\ \Eprint
  {http://arxiv.org/abs/1501.02140} {arXiv:1501.02140} \BibitemShut {NoStop}%
\bibitem [{\citenamefont {{Abbott}}\ \emph {et~al.}(2007)\citenamefont
  {{Abbott}}, \citenamefont {{Abbott}}, \citenamefont {{Adhikari}},
  \citenamefont {{Agresti}}, \citenamefont {{Ajith}}, \citenamefont {{Allen}},
  \citenamefont {{Amin}}, \citenamefont {{Anderson}}, \citenamefont
  {{Anderson}}, \citenamefont {{Araya}},\ and\ \citenamefont
  {et~al.}}]{Abbott:2007aa}%
  \BibitemOpen
  \bibfield  {author} {\bibinfo {author} {\bibfnamefont {B.}~\bibnamefont
  {{Abbott}}}, \bibinfo {author} {\bibfnamefont {R.}~\bibnamefont {{Abbott}}},
  \bibinfo {author} {\bibfnamefont {R.}~\bibnamefont {{Adhikari}}}, \bibinfo
  {author} {\bibfnamefont {J.}~\bibnamefont {{Agresti}}}, \bibinfo {author}
  {\bibfnamefont {P.}~\bibnamefont {{Ajith}}}, \bibinfo {author} {\bibfnamefont
  {B.}~\bibnamefont {{Allen}}}, \bibinfo {author} {\bibfnamefont
  {R.}~\bibnamefont {{Amin}}}, \bibinfo {author} {\bibfnamefont {S.~B.}\
  \bibnamefont {{Anderson}}}, \bibinfo {author} {\bibfnamefont {W.~G.}\
  \bibnamefont {{Anderson}}}, \bibinfo {author} {\bibfnamefont
  {M.}~\bibnamefont {{Araya}}}, \ and\ \bibinfo {author} {\bibnamefont
  {et~al.}},\ }\bibfield  {title} {\enquote {\bibinfo {title} {{Searching for a
  Stochastic Background of Gravitational Waves with the Laser Interferometer
  Gravitational-Wave Observatory}},}\ }\href {\doibase 10.1086/511329}
  {\bibfield  {journal} {\bibinfo  {journal} {\apj}\ }\textbf {\bibinfo
  {volume} {659}},\ \bibinfo {pages} {918--930} (\bibinfo {year} {2007})},\
  \Eprint {http://arxiv.org/abs/astro-ph/0608606} {astro-ph/0608606}
  \BibitemShut {NoStop}%
\bibitem [{\citenamefont {{Jenet}}\ \emph {et~al.}(2006)\citenamefont
  {{Jenet}}, \citenamefont {{Hobbs}}, \citenamefont {{van Straten}},
  \citenamefont {{Manchester}}, \citenamefont {{Bailes}}, \citenamefont
  {{Verbiest}}, \citenamefont {{Edwards}}, \citenamefont {{Hotan}},
  \citenamefont {{Sarkissian}},\ and\ \citenamefont {{Ord}}}]{Jenet:2006aa}%
  \BibitemOpen
  \bibfield  {author} {\bibinfo {author} {\bibfnamefont {F.~A.}\ \bibnamefont
  {{Jenet}}}, \bibinfo {author} {\bibfnamefont {G.~B.}\ \bibnamefont
  {{Hobbs}}}, \bibinfo {author} {\bibfnamefont {W.}~\bibnamefont {{van
  Straten}}}, \bibinfo {author} {\bibfnamefont {R.~N.}\ \bibnamefont
  {{Manchester}}}, \bibinfo {author} {\bibfnamefont {M.}~\bibnamefont
  {{Bailes}}}, \bibinfo {author} {\bibfnamefont {J.~P.~W.}\ \bibnamefont
  {{Verbiest}}}, \bibinfo {author} {\bibfnamefont {R.~T.}\ \bibnamefont
  {{Edwards}}}, \bibinfo {author} {\bibfnamefont {A.~W.}\ \bibnamefont
  {{Hotan}}}, \bibinfo {author} {\bibfnamefont {J.~M.}\ \bibnamefont
  {{Sarkissian}}}, \ and\ \bibinfo {author} {\bibfnamefont {S.~M.}\
  \bibnamefont {{Ord}}},\ }\bibfield  {title} {\enquote {\bibinfo {title}
  {{Upper Bounds on the Low-Frequency Stochastic Gravitational Wave Background
  from Pulsar Timing Observations: Current Limits and Future Prospects}},}\
  }\href {\doibase 10.1086/508702} {\bibfield  {journal} {\bibinfo  {journal}
  {\apj}\ }\textbf {\bibinfo {volume} {653}},\ \bibinfo {pages} {1571--1576}
  (\bibinfo {year} {2006})},\ \Eprint {http://arxiv.org/abs/astro-ph/0609013}
  {astro-ph/0609013} \BibitemShut {NoStop}%
\bibitem [{\citenamefont {{Ungarelli}}\ and\ \citenamefont
  {{Vecchio}}(2001)}]{Ungarelli:2001aa}%
  \BibitemOpen
  \bibfield  {author} {\bibinfo {author} {\bibfnamefont {C.}~\bibnamefont
  {{Ungarelli}}}\ and\ \bibinfo {author} {\bibfnamefont {A.}~\bibnamefont
  {{Vecchio}}},\ }\bibfield  {title} {\enquote {\bibinfo {title} {{Studying the
  anisotropy of the gravitational wave stochastic background with LISA}},}\
  }\href {\doibase 10.1103/PhysRevD.64.121501} {\bibfield  {journal} {\bibinfo
  {journal} {\prd}\ }\textbf {\bibinfo {volume} {64}},\ \bibinfo {pages}
  {121501} (\bibinfo {year} {2001})},\ \Eprint
  {http://arxiv.org/abs/astro-ph/0106538} {astro-ph/0106538} \BibitemShut
  {NoStop}%
\bibitem [{\citenamefont {{Cutler}}\ and\ \citenamefont
  {{Harms}}(2006)}]{Cutler:2006aa}%
  \BibitemOpen
  \bibfield  {author} {\bibinfo {author} {\bibfnamefont {C.}~\bibnamefont
  {{Cutler}}}\ and\ \bibinfo {author} {\bibfnamefont {J.}~\bibnamefont
  {{Harms}}},\ }\bibfield  {title} {\enquote {\bibinfo {title} {{Big Bang
  Observer and the neutron-star-binary subtraction problem}},}\ }\href
  {\doibase 10.1103/PhysRevD.73.042001} {\bibfield  {journal} {\bibinfo
  {journal} {\prd}\ }\textbf {\bibinfo {volume} {73}},\ \bibinfo {eid} {042001}
  (\bibinfo {year} {2006})},\ \Eprint {http://arxiv.org/abs/gr-qc/0511092}
  {gr-qc/0511092} \BibitemShut {NoStop}%
\bibitem [{\citenamefont {{Boyle}}\ and\ \citenamefont
  {{Steinhardt}}(2008)}]{Boyle:2008aa}%
  \BibitemOpen
  \bibfield  {author} {\bibinfo {author} {\bibfnamefont {L.~A.}\ \bibnamefont
  {{Boyle}}}\ and\ \bibinfo {author} {\bibfnamefont {P.~J.}\ \bibnamefont
  {{Steinhardt}}},\ }\bibfield  {title} {\enquote {\bibinfo {title} {{Probing
  the early universe with inflationary gravitational waves}},}\ }\href
  {\doibase 10.1103/PhysRevD.77.063504} {\bibfield  {journal} {\bibinfo
  {journal} {\prd}\ }\textbf {\bibinfo {volume} {77}},\ \bibinfo {eid} {063504}
  (\bibinfo {year} {2008})},\ \Eprint {http://arxiv.org/abs/astro-ph/0512014}
  {astro-ph/0512014} \BibitemShut {NoStop}%
\bibitem [{\citenamefont {{Kuroyanagi}}\ \emph {et~al.}(2014)\citenamefont
  {{Kuroyanagi}}, \citenamefont {{Tsujikawa}}, \citenamefont {{Chiba}},\ and\
  \citenamefont {{Sugiyama}}}]{Kuroyanagi:2014aa}%
  \BibitemOpen
  \bibfield  {author} {\bibinfo {author} {\bibfnamefont {S.}~\bibnamefont
  {{Kuroyanagi}}}, \bibinfo {author} {\bibfnamefont {S.}~\bibnamefont
  {{Tsujikawa}}}, \bibinfo {author} {\bibfnamefont {T.}~\bibnamefont
  {{Chiba}}}, \ and\ \bibinfo {author} {\bibfnamefont {N.}~\bibnamefont
  {{Sugiyama}}},\ }\bibfield  {title} {\enquote {\bibinfo {title}
  {{Implications of the B-mode polarization measurement for direct detection of
  inflationary gravitational waves}},}\ }\href {\doibase
  10.1103/PhysRevD.90.063513} {\bibfield  {journal} {\bibinfo  {journal}
  {\prd}\ }\textbf {\bibinfo {volume} {90}},\ \bibinfo {eid} {063513} (\bibinfo
  {year} {2014})},\ \Eprint {http://arxiv.org/abs/1406.1369} {arXiv:1406.1369}
  \BibitemShut {NoStop}%
\bibitem [{\citenamefont {{Sakakihara}}\ and\ \citenamefont
  {{Soda}}(2015)}]{Sakakihara:2015aa}%
  \BibitemOpen
  \bibfield  {author} {\bibinfo {author} {\bibfnamefont {Y.}~\bibnamefont
  {{Sakakihara}}}\ and\ \bibinfo {author} {\bibfnamefont {J.}~\bibnamefont
  {{Soda}}},\ }\bibfield  {title} {\enquote {\bibinfo {title} {{Primordial
  Gravitational Waves in Bimetric Gravity}},}\ }\href@noop {} {\bibfield
  {journal} {\bibinfo  {journal} {ArXiv e-prints}\ } (\bibinfo {year}
  {2015})},\ \Eprint {http://arxiv.org/abs/1504.04969} {arXiv:1504.04969
  [hep-th]} \BibitemShut {NoStop}%
\bibitem [{\citenamefont {Vainshtein}(1972)}]{Vainshtein:1972sx}%
  \BibitemOpen
  \bibfield  {author} {\bibinfo {author} {\bibfnamefont {A.I.}\ \bibnamefont
  {Vainshtein}},\ }\bibfield  {title} {\enquote {\bibinfo {title} {{To the
  problem of nonvanishing gravitation mass}},}\ }\href {\doibase
  10.1016/0370-2693(72)90147-5} {\bibfield  {journal} {\bibinfo  {journal}
  {Phys.Lett.}\ }\textbf {\bibinfo {volume} {B39}},\ \bibinfo {pages}
  {393--394} (\bibinfo {year} {1972})}\BibitemShut {NoStop}%
\bibitem [{\citenamefont {{de Rham}}\ \emph {et~al.}(2013)\citenamefont {{de
  Rham}}, \citenamefont {{Heisenberg}},\ and\ \citenamefont
  {{Ribeiro}}}]{de-Rham:2013ab}%
  \BibitemOpen
  \bibfield  {author} {\bibinfo {author} {\bibfnamefont {C.}~\bibnamefont {{de
  Rham}}}, \bibinfo {author} {\bibfnamefont {L.}~\bibnamefont {{Heisenberg}}},
  \ and\ \bibinfo {author} {\bibfnamefont {R.~H.}\ \bibnamefont {{Ribeiro}}},\
  }\bibfield  {title} {\enquote {\bibinfo {title} {{Quantum corrections in
  massive gravity}},}\ }\href {\doibase 10.1103/PhysRevD.88.084058} {\bibfield
  {journal} {\bibinfo  {journal} {\prd}\ }\textbf {\bibinfo {volume} {88}},\
  \bibinfo {eid} {084058} (\bibinfo {year} {2013})},\ \Eprint
  {http://arxiv.org/abs/1307.7169} {arXiv:1307.7169 [hep-th]} \BibitemShut
  {NoStop}%
\bibitem [{\citenamefont {{Heisenberg}}(2014)}]{Heisenberg:2014aa}%
  \BibitemOpen
  \bibfield  {author} {\bibinfo {author} {\bibfnamefont {L.}~\bibnamefont
  {{Heisenberg}}},\ }\bibfield  {title} {\enquote {\bibinfo {title} {{Quantum
  corrections in massive bigravity and new effective composite metrics}},}\
  }\href@noop {} {\bibfield  {journal} {\bibinfo  {journal} {ArXiv e-prints}\ }
  (\bibinfo {year} {2014})},\ \Eprint {http://arxiv.org/abs/1410.4239}
  {arXiv:1410.4239 [hep-th]} \BibitemShut {NoStop}%
\end{thebibliography}%

\end{document}